\documentclass[12pt,preprint]{aastex}

\shorttitle{GRB Pulse Unification}
\shortauthors{Hakkila and Preece}

\begin{document}

\title{Unification of Pulses in Long and Short Gamma-Ray Bursts: \\ Evidence from Pulse Properties and their Correlations}

\author{Jon Hakkila\altaffilmark{1} and Robert D. Preece\altaffilmark{2}} 

\affil{$^1$Dept. Physics and Astronomy, The College of Charleston, Charleston, SC  29424-0001\\$^2$Department of Physics and Astronomy, University of Alabama in Huntsville, Huntsville, AL 35899\\}
\email{hakkilaj@cofc.edu}

\begin{abstract}

We demonstrate that distinguishable gamma-ray burst pulses exhibit similar behaviors as evidenced by correlations among the observable pulse properties of duration, peak luminosity, fluence, spectral hardness, energy-dependent lag, and asymmetry. Long and Short burst pulses exhibit these behaviors, suggesting that a similar process is responsible for producing all GRB pulses. That these properties correlate in the observer's frame indicates that intrinsic correlations are strong enough to not be diluted into insignificance by the dispersion in distances and redshift. We show how all correlated pulse characteristics can be explained by hard-to-soft pulse evolution, and we demonstrate that ``intensity tracking'' pulses not having these properties are not single pulses; they instead appear to be composed of two or more overlapping hard-to-soft pulses. In order to better understand pulse characteristics, we recognize that hard-to-soft evolution provides a more accurate definition of a pulse than its intensity variation. This realization, coupled with the observation that pulses begin near-simultaneously across a wide range of energies, leads us to conclude that the observed pulse emission represents the energy decay resulting from an initial injection, and that one simple and as yet unspecified physical mechanism is likely to be responsible for all gamma-ray burst pulses regardless of the environment in which they form and, if GRBs originate from different progenitors, then of the progenitors that supply them with energy.

\end{abstract}




\section{Introduction}

Gamma-ray burst (GRB) prompt emission is generally representative of coherent, {\em pulsed} radiation, rather than of a random walk, a white noise process, or some other type of noise process. In recent years, pulse properties have provided increasingly valuable constraints on the physics responsible for GRB prompt emission. These properties have included (1) temporal asymmetry characterized by longer decay than rise rates, (2) hard-to-soft spectral evolution, and (3) broadening at lower energies (e.g. \cite{gol83, nor96, lia96, rrm00, nor02, ryd05}). 

The vast majority of GRB pulses have similar shapes; at least they are similar enough that most can be fitted by a simple four-parameter empirical pulse model \citep{nor05}. The model has allowed many isolated and semi-overlapping GRB pulses to be extracted \citep{nor05,hak08,hak09a}. However, unique extraction of heavily-overlapping pulses and very low signal-to-noise pulses is difficult \citep{hak09a}, and the characteristics of these unfitted pulses are unknown due to the difficulty in making measurements as well as the inherent non-uniqueness of the solutions (e.g. \cite{act70}).

When applied to pulse data contained in individual energy channels, the pulse fit model demonstrates not only that pulses have longer durations at lower energies, but that they also peak later at these lower energies \citep{hak08, hak09a}. This energy-dependent pulse lag is different for each pulse (this can be verified using either the {\em pulse peak lag} \citep{hak08, hak09a} or by applying the cross-correlation function \citep{ban97} to pulses, indicating that greater information content resides in the pulse lags than can be obtained from measuring a single lag over the duration of the GRB. The pulse peak lag inversely correlates with the pulse peak luminosity \citep{hak08}, indicating that this relation is the basis of the lag vs.\ luminosity relation obtained for integrated prompt emission \citep{nor00}. Furthermore, pulse duration also correlates directly with pulse peak luminosity, allowing easily-measured pulse durations to be used as a GRB luminosity indicator \citep{hak08}. In addition to BATSE data, this technique has recently been applied to HETE-2 \citep{ari10} and Swift data ({e.g. \cite{chi10,mag10}), and the technique appears to provide a promising way of estimating GRB redshifts \citep{hak09b}.

In recent preliminary work, \cite{hak09a} have demonstrated that other pulse spectral and temporal characteristics (besides lag and duration) correlate with pulse luminosity, and thus with each other. Pulse duration correlates with pulse asymmetry and pulse fluence and anti-correlates with pulse hardness and pulse peak flux; all of these measures are thus pulse peak luminosity indicators. \cite{hak09c} have further shown that, within measurement uncertainties, pulse emission begins simultaneously at all energies.

The ubiquitous presence of GRB pulses, along with their correlative properties, redirects attention from bulk GRB emission towards its constituent structures. Many luminosity-correlating properties attributed to bulk GRB prompt emission are actually properties of convolved pulses, and can be understood as such \citep{hak09a}. As previously mentioned, a GRB's lag \citep{nor00} is a heterogeneous composite of the individual pulse lags favoring the high intensity, short duration pulses \citep{hak08, hak09a, ari10}. Variability \citep{rei01}, measuring the changing luminosity of a GRB, increases and decreases with each pulse, and thus indicates a GRB's multi-pulse structure. The peak energy $E_{\rm peak}$ (the peak of the $\nu F_{\nu}$ spectrum) correlates with the isotropic energy $E_{\rm iso}$ \citep{ama02}, yet $E_{\rm peak}$ is the time-integrated peak of the combined spectra of all pulses while $E_{\rm iso}$ is the summed fluence luminosity of all pulses. The Internal Luminosity Function (ILF), or $\psi(L)$ \citep{hor97}, is the observed luminosity distribution within a GRB, and the quantity $\psi(L) \Delta L$ represents the fraction of time that the GRB's luminosity is found between $L$ and $L+\Delta L$. The ILF summarizes information about the distribution of luminosity within a burst without explicitly describing the order in which the emission occurred, or if the emission was pulsed. However, the ILF measures a convolution of the luminosity within each pulse along with the number of pulses which contribute to the total emission \citep{hak07}. The correlation between GRB peak luminosities and $\psi(L)$ thus indicates that GRB luminosity is driven by pulse structure rather than being a generic property of the prompt emission. 

Since GRBs can be classified into Long and Short bursts on the basis of duration, hardness, and fluence (e.g. \cite{kou93, muk98, hak03}), the preliminary \cite{hak09a} results demonstrate that the correlations between pulse properties are common to pulses originating in both the Short and Long GRB classes. Furthermore, there is substantial overlap between the properties of the shortest pulses in Long GRBs and Short GRB pulses, suggesting the existence of a single process.

In this manuscript, we demonstrate that correlative GRB pulse properties are universal for Short and Long GRBs, and are thus progenitor-independent (if Short and Long GRBs originate from different progenitors). We further demonstrate that these simple pulse behaviors place strong constraints on GRB physics.

\section{Measuring Pulse Properties}

In this study, we analyze GRBs cataloged by the Burst And Transient Source Experiment (BATSE) on NASA's Compton Gamma-Ray Observatory (CGRO). By studying GRB pulses obtained from a single experiment, we limit our sample to a homogeneous one with well-understood instrumental and sampling biases. Inclusion of GRBs detected by other instruments (e.g. HETE-2, Swift, and Fermi) introduces a wide range of poorly-understood and pervasive systematic biases that complicate the study of GRBs across instruments: count rates from BATSE cannot be directly compared to those of other instruments because each instrument is characterized by a different spectral range, temporal response, and energy-dependent sensitivity. For example, (1) Swift and HETE-2 are not sensitive at the higher energies of BATSE and Fermi, where the GRB photon count rates are highest, and (2) BATSE has a larger surface area and therefore a lower photon detection threshold than the similar detectors on Fermi's GBM experiment. BATSE's well-understood systematics have been extensively documented, which can be used to determine the viability of extracted pulse properties.

The pulses described here are part of a BATSE GRB pulse catalog currently being produced (Hakkila et al., in preparation) by an analysis of the Current BATSE Catalog (\url{http://www.batse.msfc.nasa.gov/batse/grb/catalog/current/}). The process involves extracting pulses from summed 4-channel 64 ms GRB data using the Bayesian Blocks methodology \citep{sca98} and subsequently fitting the possible pulse time intervals with the four-parameter model of \cite{nor05}. 

The Bayesian Blocks process \citep{sca98} at the root of this analysis is a statistical model by which a photon-counting light curve is segmented into a sequence of discontinuous constant levels; each level indicates that the background is statistically different than the segments preceding and following it. The result of a Bayesian Block segmentation looks like histogram with variable-width bins. Segments with elevated backgrounds are regions that possibly contain pulses.

We assume that candidate pulses can be modeled with the empirical functional pulse form of \cite{nor05}:
\begin{equation}
I(t) = A \lambda \exp^{[-\tau_1/(t - t_s) - (t - t_s)/\tau_2]},
\end{equation}
where $t$ is time since trigger, $A$ is the pulse amplitude, $t_s$ is the pulse start time, 
$\tau_1$ and $\tau_2$ are characteristics of the pulse rise and pulse decay, and the constant 
$\lambda = \exp{[2 (\tau_1/\tau_2)^{1/2}]}$. The 4-channel pulse peak time occurs at time $\tau_{\rm peak} = t_s + \sqrt{\tau_1 \tau_2}$. As each segment is searched for a pulse, statistically-insignificant intervals are merged into the surrounding intervals until an optimal set of model pulses is identified for the GRB. The criterion for identifying a statistically-significant pulse is based on the idea of a {\em dual timescale peak flux} \citep{hak03}, which does not bias long, faint pulses relative to short, bright ones. The long timescale peak flux is based on the 4-channel fluence (time-integrated photon counting flux $S_{\rm tot}$) divided by the pulse duration $w$ while the short timescale is based on the 4-channel 64 ms peak flux ($p_{\rm 64}$).

The criterion generally used in our analysis to select pulses is
\begin{equation}
p_{\rm 64} S_{\rm tot}/w > n^2 b
\end{equation}
where b is the average background counts in a 64-ms interval, $w$ is the pulse duration (described below), and $n$ is the number of standard deviations above background that the pulse peak needs to be in order to be considered. The default setting of the automated code is $n=3$, which works well for most isolated pulses and many partially overlapping pulses. However, operator assistance is needed when inspection indicates that distinct pulses have not been properly fitted. For example, thresholds as low as $n=1$ have been required to fit some long, faint, isolated pulses for which the BATSE background rate is particularly well-behaved. Larger values of $n$ (e.g. $n>10$) are needed to extract some overlapping pulses in complex GRBs because the greatest source of noise to fitting a pulse can be the local background caused by overlapping pulses.

Four examples of this process are demonstrated in Figure 1; single-pulsed BATSE triggers 1221 and 432, and multi-pulsed triggers 1443 and 1468. Each Bayesian block segment is identified by vertical lines which overlay the raw counts data. The pulses fitted to the data are also shown, as is the overall fit. 

Pulses fitted in the summed 4-channel data are subsequently extracted from individual energy channels using the 4-channel pulse parameters as starting points. Although the process is fairly robust, low intensity pulses, pulses that strongly overlap, and/or very short pulses indistinguishable from Poisson noise are not considered to have been successfully fitted, and are excluded from this discussion. 

A variety of model-dependent, observable pulse properties are extracted from the multichannel parameters $t_s$, $A$, $\tau_1$, and $\tau_2$. These observables are extracted independently for energy channels 1, 2, 3, and 4 as well as for the summed BATSE 4-channel data, although low signal-to-noise makes channel 4 pulse properties very difficult to measure. These observables (chosen so that they can be compared to properties of the prompt emission) include pulse durations, pulse peak lags, pulse peak fluxes, pulse asymmetries, and pulse spectral hardnesses. 

The {\em pulse duration} $w$ is obtained from the summed four-channel data and is defined as $w = [9 + 12\sqrt{\tau_1/\tau_2}]^{1/2}$; this is the interval between times when the pulse amplitude is $A {e^{-3}}$ (note that \cite{nor05} uses the shorter interval between times when the pulse amplitude is $A {e^{-1}}$). {\em Pulse peak lags} $l$ are the differences between the pulse peak times in different energy channels (pulse peak times are given by $\tau_{\rm peak} = t_s + \sqrt{\tau_1 \tau_2}$). Pulse peak lags can be obtained for any pulse between two energy channels, although we define the standard pulse peak lag $l$ as that measured between energies of 100 to 300 keV (BATSE channel 3) and 25 to 50 keV (BATSE channel 1). Pulse durations and pulse peak lags are measured in units of seconds. 

The {\em pulse peak flux} $p_{256}$ is the peak flux (in counts $\rm cm^{-2} s^{-1}$) of the fitted pulse defined on the 256 ms timescale. The {\em pulse fluence} ($S$) is the time-integrated flux summed over the pulse duration in the 25 to 300 keV range (channels 1 through 3), given in units of ergs $\rm cm^{-2}$. We define the fluence in this way (rather than in counts $\rm cm^{-2}$) so that the results can be more directly compared to published BATSE burst fluences. In order to make this conversion, we have assumed an approximate average energy for each BATSE channel, and we have also excluded channel 4 fluence because of BATSE's lower quality instrumental response at these energies.

The pulse asymmetry $\kappa$ is defined as $\kappa = w/(3 + 2\sqrt{\tau_1/\tau_2}$); it ranges from a value of $\kappa=0$ for a symmetric pulse to $\kappa=1$ for an asymmetric pulse with a rapid rise and slow decay.  {\em Hardness ratios} HR are constructed by dividing pulse fluences in two different energy channels; we use the hardness ratio HR$=S_3/S_1$, where $S_3$ is the channel 3 energy fluence and $S_1$ is the channel 1 energy fluence. Both the pulse asymmetry and the hardness ratio are unitless parameters.

The GRBs have been classified using the BATSE-dependent data mining definition of \cite{hak03}, which classifies GRBs as Short if they satisfy the inequality (T90 $<1.954$) OR ($1.954 \le T90 < 4.672$ AND ${\rm HR}_{321} > 3.01$), where ${\rm HR}_{321} = S_3/(S_2 + S_1)$ \citep{muk98}, and where $S_2$ is the channel 2 fluence. If GRBs do not satisfy the aforementioned inequality, then they are classified as Long.

Electronic Table 1 contains the GRB pulse properties used in this study, along with their uncertainties. This table contains preliminary data from the BATSE pulse catalog currently under development. Current pulse extraction has yielded to date 1338 pulses in 610 BATSE GRBs.

\section{Pulse Property Correlations}

The properties of the GRB pulses examined in this study exhibit many strong correlations; these can be seen in mutipanel Figure 1. This figure demonstrates the relationships between the observable pulse properties of duration, lag, peak flux, hardness ratio, asymmetry, and fluence. Pulses from the Short class of GRBs are shown in blue, while pulses from the Long GRB class are denoted in red. Many correlations and anti-correlations are apparent by inspection, and the statistical significances of the correlations are summarized in Table 2. Since Short GRB pulses have short durations and large amplitudes, they occupy an extreme position in each distribution. The correlations and anti-correlations illustrated here have been described previously in \cite{hak08} and \cite{hak09a}, but are much more statistically significant given the larger size of this pulse sample.

A strong correlation exists between pulse duration and pulse peak lag; this correlation is present for both Long and Short burst pulses. However, not all pulse lags are positive; one-third are zero or negative. Most of the negative lags are consistent with their measurement uncertainties (indicating that they are simply too small to accurately measure), although large negative lag values are suggestive of systematic extraction issues involving unresolved overlapping pulses (e.g. see \cite{hak09a}). Figure 2 demonstrates the relationship between pulse duration and pulse peak lag for the pulses with positive lags. Assuming that pulses have positive lags, pulse peak lag appears to be a proxy for hard-to-soft pulse evolution. By this we mean that the timescale for pulse emission correlates with the timescale of the hard-to-soft evolution. This effect is not distorted by cosmological redshift $z$ to first order, because the values of both duration and lag in the emitted frame are $(1+z)^{-1}$ the values measured in the observer's frame (in making this statement, we purposefully ignore effects caused by the $k$-correction and the spectral range of the instrument, which will be discussed later).

Pulse duration anti-correlates with pulse peak flux for both Long and Short burst pulses. The brightest pulses (as measured by peak flux) found within observed GRBs are also the shortest pulses. This is surprising given the strong correlation between pulse peak luminosity and pulse duration (e.g. \cite{hak09a,ari10,chi10}), because it indicates the correlations are strong enough to not be diluted into insignificance by the dispersion in GRB distances. The photon peak flux on the proper motion distance $d_M$ and redshift $z$ is proportional to $[(1+z)d_M^2]^{-1}$. Again, the first-order effects of time dilation cancel in the comparison between duration and peak flux, so that only the inverse square law disperses the relation. However, the similarities between the peak flux vs. duration distribution and the luminosity vs. duration distribution suggest that the peak flux distribution of GRB pulses is closely related to the pulse peak luminosity function. We note that the pulses from Short GRBs appear to extend the correlation found for Long burst pulses such that the shortest pulses from Long GRBs overlap the Short GRB distribution.

Pulse spectral hardness is anti-correlated with pulse duration such that shorter, brighter pulses are generally harder than longer, fainter pulses. The relationship is weaker for Short GRB pulses than for Long ones, although it still seems significant. We note that spectral hardness has no $z$-dependence, but it does require a $k$-correction, which may be significant for bursts that differ in redshift by factors of order 2. Since duration has not $z$-dependence, there is a scatter in the correlation caused by the redshift dependence of the duration relative to the spectral hardness.

Pulse asymmetry also correlates with pulse duration, such that the most asymmetric pulses are long duration, long lag, low peak flux, soft pulses. Since pulse shape is difficult to measure for pulses having durations approaching the 64 ms timescale (Poisson background fluctuations can significantly mask pulse characteristics), Short GRB pulses tend to be fitted with a bimodal asymmetry distribution with asymmetries approaching the extreme values of either $\kappa=0$ or $\kappa=1$, and show no statistically significant correlation with pulse duration. Pulse asymmetry also has no $z$-dependence, so correlations between it and other pulse properties is dispersed in those properties by a factor of $(1+z)^{-1}$.

Pulse energy fluence $S$ is strongly correlated with pulse duration; longer pulses contain more energy than short pulses.  A strong anti-correlation also exists between pulse peak flux and pulse fluence, indicating that the pulses that emit their energies in the shortest time interval generally produce the fewest photons over their duration. And, to a first order approximation, the harder spectrum of a short pulse does not provide enough energy to compensate for the larger number of photons produced by a longer pulse. Thus, the pulses with the largest energy fluence are long, soft, low peak flux pulses. Because we have chosen to measure fluence in energy units, the redshift dependence is again proportional to $[(1+z)d_M^2]^{-1}$ (e.g. see also \cite{mes11}). Thus, redshift only causes a dispersion in $S$ when compared to $\kappa$ and HR, which have no redshift dependence. However, like peak flux, the inverse square law introduces a $d_M^{-2}$ dispersion when being compared to all pulse properties other than peak flux.

Most of the remaining relationships between GRB pulse properties described in Table 2 are as expected, given the aforementioned correlations and anti-correlations (the exception is the hardness vs. fluence relation, which will be discussed in section 5). The difficulty in accurately measuring lag and asymmetry for very short pulses makes it hard to accurately measure correlations between these properties and other ones for short duration GRB pulses. 

We note that, since lags and durations of Long GRB pulses are indicators of pulse peak luminosity, the observer frame attributes of spectral hardness, pulse asymmetry, fluence, and peak flux are also luminosity indicators. We also note that pulses from Long and Short GRBs appear to represent a continuum of values; we will return to this point in section 8.

\section{Intra-Burst Pulse Properties}

The previous discussion refers to correlations between all pulses, regardless of the bursts in which they originate (e.g.\ {\em inter-burst pulse} correlations).  We can also look at the distributional properties of pulses within GRBs (e.g.\ {\em intra-burst pulse} properties). Although we have only had limited success fitting pulses in complex GRBs, we have many bursts with two, three, and four pulses in them. However, since only a small number of Short GRBs have multiple distinct pulses on the 64-ms timescale, we treat multi-pulsed Short GRBs as if they are part of the same sample as the multi-pulsed Long GRBs.

Averaging together the properties of pulses within multi-pulsed GRBs gives us some sense of the bulk intra-burst pulse distribution. Figures 4 through 7 indicate the mean values of several key, easily-measured pulse properties measured within individual GRBs: duration, peak flux, hardness, and fluence (the error bars indicate the mean error for each averaged measurement). The {\em mean} pulse properties for multi-pulsed GRBs correlate similarly to the bulk properties of individual pulses. In other words, average pulse duration is apparently a good indicator of average pulse brightness, hardness, lag, and fluence, just as is found in single-pulsed bursts (although the variation of pulse asymmetries, coupled with large measurement uncertainties, make them an uninformative average pulse property).

It is more difficult to ascertain if the pulses within multi-pulsed GRBs themselves obey the correlations identified in Figure 2 and Table 2. Overlapping pulses in multi-pulsed bursts are the most difficult to fit and typically provide the least reliable pulse property measurements. In other words, bursts with the largest numbers of pulses tend to have the smallest numbers of accurately fitted pulses. Thus, in attempting to verify the correlative relationships between pulses within bursts, we will be limited to intra-pulse data consisting of bursts with a small number of pulses having well-measured properties along with bursts having many pulses but with less reliable pulse property measurements.

As a simple test of whether the correlations exist, we calculate the linear gradients (slopes) of pulse parameters within multi-pulse GRBs. We apply least squares to the properties of all pulses fitted within each multi-pulsed GRB to learn more about the intra-burst pulse distribution. The resultant gradient of each pulse property relative to the others indicates whether the properties within the burst are correlated (positive) or anti-correlated (negative). The percentage of multi-pulsed GRBs exhibiting correlated behaviors is then compared to the results of the inter-burst pulse duration shown in Figure 2 and Table 2 and is demonstrated in Table 3; if somewhat more than 50\% of the multi-pulsed GRBs have the same general gradient as that obtained from the inter-burst pulse distribution then the two distributions have the same general behavior. Since these percentages average 62\%, the behaviors of the intra-burst pulse distribution have the same sense but otherwise appear to be generally weaker than the power-law forms of the inter-burst pulse distribution. Although these weaker relationships might indicate weaker intrinsic relationships between pulse properties within a burst, they might simple reflect a greater difficulty in accurately measuring these properties in complex GRBs.

\section{Possible Contributions of Sampling Biases to Pulse Property Correlations}

The observed pulse correlations are not caused by the way pulses have been extracted, or by other selection biases (e.g. \cite{hig96,mee00}). Figure 8 shows that the peak flux vs.\ duration correlation is strongly present even for single-pulsed GRBs triggering on the 256 ms timescale (from the BATSE $C_{\rm max}/C_{\rm min}$ Table found at {\url{http://www.batse.msfc.nasa.gov/batse/grb/catalog/current/}). All of the pulses included in Figure 8 have been selected on the basis of peak flux rather than the dual-timescale pulse trigger used in the selection of pulses from multi-pulsed bursts.  The sample is lacking in short pulses that trigger preferentially on the 64 ms timescale (and at the cost of long duration, low peak flux pulses) and in long pulses that trigger preferentially on the 1024 ms timescale (at the cost of short duration pulses), yet still shows a pronounced correlation between peak flux and duration. The overall sample shown in Figure 2 has an additional advantage of triggering on BATSE's three different trigger timescales, since this sample includes pulses that triggered on 64 ms OR 256 ms OR 1024 ms. The multi-timescale trigger of the entire sample is thus less susceptible to selection biases than a catalog selected solely on the basis of a single timescale trigger would be. As the remaining pulses in the pulse catalog have also been selected using a dual-timescale trigger (peak flux vs.\ fluence), it appears that biases related to the trigger are not responsible for the existence of the pulse peak flux vs.\ duration correlation.

We can also determine if the dual timescale peak flux used to select pulses is somehow responsible for creating the observed pulse property correlations. We have estimated the threshold dual timescale peak flux limits of extracted GRB pulses using equation 2 in conjunction with mean BATSE background counts; the results are shown in Figure 9 in terms of the two constituent components of the dual timescale peak flux: the counts fluence divided by the duration, and the 64 ms peak flux. The solid line represents the standard dual timescale peak flux threshold described in equation 2 when $b=3$, while the dotted line represents the near minimum threshold when $b=1$. The approximate durations of the plotted pulses are indicated on the diagram, since pulse duration correlates strongly with fluence. The diagram suggests that background is not an issue for pulses with durations shorter than $\approx 10$ s, although it makes it increasingly difficult to recognize and extract pulses longer than 100 s. Thus, because the slope of the data is roughly perpendicular to the pulse-fitting threshold, the threshold is not responsible for the strong pulse property correlations observed in short, bright pulses. Furthermore, correlations continue to adequately describe peak fluxes and fluences of pulses fainter than this threshhold, suggesting that the correlations also hold for longer, fainter pulses. Although we have demonstrated that trigger biases are not responsible for the pulse correlations, it does appear that they prevent us from detecting very long, low-amplitude, single-pulsed bursts (e.g. pulses on the order of hundreds to thousands of seconds).

The analysis has some other important selection biases as well. As mentioned before, the analysis purposefully excludes bright pulses which cannot be uniquely identified due to pulse overlap. Similarly, pulses that are too faint and/or too short (if there are any of these) will be excluded on the basis that they can be confused with Poisson noise. However, the evidence indicates there does not seem to be a systematic exclusion of pulses capable of explaining the observed pulse correlations.


The one statistically-significant correlation that represents a difference between Long and Short GRB pulses is that between hardness and fluence (see Table 2). Short GRB pulses exhibit a strong correlation between hardness and fluence, whereas Long GRB pulses exhibit a moderately strong anti-correlation between these properties. This disparity is not real, but is due the fluence definition being used in this analysis. Short duration pulses, and particularly those belonging to the Short GRB class, are very hard and often have a significant amount of channel 4 fluence. Since our fluence definition excludes channel 4 (because most pulses are too difficult to accurately fit at these energies), and since the average energy per photon is largest in channel 4, we are systematically underestimating the energy contribution of the shortest, hardest pulses. The shorter and brighter a pulse is, the more our fluence definition has underestimated the high energy contribution.

As has been mentioned, the effect of redshift cancels in almost all the aforementioned pulse property correlations to first order. However, energy-dependent corrections resulting from the instrumental bandwidth ($k$-corrections) are also important, although more difficult to assess. Pulses observed at large distances should have their high energy emission shifted into low-energy channels and their low-energy emission shifted out of the instrumental passband.  Increased distance will thus cause $E_{\rm peak}$ to shift to lower energy channels, sometimes moving it out of the instrumental passband. Most distant pulses will appear softer as $E_{\rm peak}$ is shifted to lower energies. Since pulses have shorter durations at high energies than at low energies, highly-redshifted pulses should appear to be shorter than their nearby counterparts; this effect works in the opposite sense of time dilation. Instruments that are insensitive to high energy emission, such as Swift and HETE-2, cannot generally detect the high energy pulse peaks occurring early in the pulse emission; pulses from these experiments thus often appear to be less peaked and more time-symmetric than their BATSE counterparts.

\section{Pulse Evolution}

As evidenced in Table 2, the correlated characteristics of GRB pulse emission are all related to the pulse spectral lag, which measures hard-to-soft pulse evolution. The pulse peak lag, denoting the pulse peak delay between BATSE channel 3 and BATSE channel 1, is only one measure of this {\em internal pulse evolution}. Pulse peak lags also exist between all BATSE energy channels, reflecting the fact that the pulse intensity peaks first in the highest energy channel and subsequently in progressively lower energy channels. Furthermore, pulses start near-simultaneously in all BATSE energy channels \citep{hak09c}; when coupled with the fact that pulses peak earlier at high energies, this indicates that hard emission dominates over soft emission early in the pulse, and decays away faster to be replaced by soft emission later in the pulse \citep{ric96}. This evolution is demonstrated in Figure 10 for single-pulsed BATSE Trigger 1883.

There are other techniques for identifying pulse evolution. Perhaps the most direct way is via the time variability of the $\nu F_\nu$ pulse spectrum via the Band spectral model \citep{ban93}. Three characteristics are used to parameterize the spectral shape: the low-energy power-law index $\alpha$, the high-energy power-law index $\beta$, and the peak of the spectral energy distribution $E_{\rm peak}$. Pulse spectral softening indicates a systematic evolution of these three spectral parameters, and primarily a shift from high $E_{\rm peak}$ to low $E_{\rm peak}$.

Spectra of hard-to-soft (classic FRED, denoting Fast Rise Exponential Decay) pulses have been observed to undergo a decay of $E_{\rm peak}$ throughout the pulse (e.g.\ \cite{cri99, koc03, kan06, pen09a}). The hardest spectra are observed early in pulse evolution, with the highest $E_{\rm peak}$ values occurring when pulse flux is just beginning to rise. Pulses undergo a time decay of $E_{\rm peak}$ that appears to be exponential, but which can be fit by a linear function in many cases \citep{cri99}. Interestingly, there is nothing special about the $E_{\rm peak}$ at the time of the pulse peak flux; there is no change in the spectral decay that delineates the pulse peak from any other time during pulse spectral evolution.

Although hard-to-soft evolution is the most common pulse evolution (e.g. \cite{cri99}, {\em intensity tracking} pulse evolution has been identified in a smaller number of pulses (e.g. \cite{for95, lia96, kan06, pen09b,lu10}). In intensity tracking evolution, spectral hardness is found to correlate with pulse flux. 

Intensity tracking pulse evolution appears to be inconsistent with pulse spectral lag measurements, since it cannot result in lags that extend across multiple energy channels. To better understand intensity tracking pulses, we select a sample of 22 hard-to-soft and intensity tracking pulses common to our BATSE pulse database from recent analyses by \cite{pen09a, pen09b, lu10}. 

In Table 3 we summarize the characteristics of these pulses and the GRBs from which they are extracted. Six pulses must be excluded from our analysis, as they have been classified as both intensity tracking and hard-to-soft in various studies, leaving eleven hard-to-soft and five intensity tracking pulses. Three of the intensity tracking and two of the hard-to-soft pulses are discarded because they have been fitted in our pulse catalog by two or more overlapping pulses. Moreover, all but one of the pulses classified as intensity tracking are found to originate in complex, multi-pulsed GRBs, where overlapping pulses are common. Hard-to-soft pulses are more often found as isolated pulses. We discard these ambiguous pulses from our sample, reducing it to nine hard-to-soft and two intensity tracking pulses that satisfy our criterion in equation 2. Four of the pulses (intensity tracking pulse 1733 and hard-to-soft pulses 2662 and 2665) have bumps prior to and/or after the pulse peak strongly suggesting the presence of additional long duration, low amplitude pulses. The remaining intensity tracking pulse (the first pulse in complex BATSE trigger 2812) appears to have two peaks that are most readily apparent in channels 2 and 3. Reconsidering these four pulses on the basis that they are potentially ambiguous, our sample contains only seven hard-to-soft pulses and no intensity tracking pulses.

Although the evidence preliminarily suggests that intensity tracking pulses are composed of overlapping hard-to-soft pulses, we re-examine some of our excluded pulses to see if there is some characteristic that they have in common. Indeed, one of the intensity tracking pulses and two of the pulses ambiguously classified have similar long duration, time-symmetric light curves. These pulses all have large non-Poisson scatter occurring near-simultaneously in several energy channels, which is consistent with the characteristics defining intensity tracking pulses. However, the instantaneous $E_{\rm peak}$ of these pulses starts high at the pulse onset before it begins to track the flux, and $E_{\rm peak}$ resets each time that non-Poisson scatter occurs. In other words, Occam's razor suggests that each of these pulses is actually composed of overlapping, short duration, hard pulses that are themselves consistent with the aforementioned correlated pulse properties. Thus, intensity tracking pulse behavior is unlikely, and simple hard-to-soft pulse behavior appears to be the dominant and perhaps only mode of pulse evolution.

The presence of faint, unresolved pulses provides a possible explanation for many ``outliers'' shown in Figures 2 and 3; despite our precautions, pulses exist exhibiting large dispersions from the otherwise strong correlative relationships. We believe that these are likely to be unresolved, merged pulses. Preliminary Monte Carlo analysis suggests that it is difficult to resolve two similar pulses peaking near-simultaneously, and that the characteristics taken on by the merged pulse are not simply a sum of the constituent pulse characteristics.

Generally, these results suggest that the empirical four-parameter pulse model of \cite{nor05} can adequately fit the vast majority of resolved, non-overapping pulses. 

\section{Peak Luminosity: a Consequence of Pulse Evolution}

Pulse duration and lag strongly anti-correlate with pulse peak luminosity. They also anti-correlate with pulse peak flux. The relationship between pulse peak flux and pulse duration is similar to, but has a broader spread than the relationship between pulse peak luminosity and pulse duration (compare Figure 2 with \cite{hak08} Figure 4). This is somewhat surprising, since the pulse duration vs.\ peak luminosity relation was established using GRBs with known redshift and corrected for luminosity distance, the inverse square law, and cosmological time dilation. That anti-correlations between pulse peak flux and duration are apparent without making these corrections (and without even converting photon flux to energy flux) indicates that dispersive effects caused by the inverse square law and relativistic cosmology do not cause enough observational scatter to destroy the intrinsic relationships. This is contrary to common understanding, as relativistic cosmology was once thought to be of central importance in the understanding of observed GRB characteristics (see the review by \cite{pir05}).  

It is perhaps not surprising that intrinsic pulse characteristics are so much stronger than relativistic cosmological effects given the presumed relativistic ejecta origins of GRB prompt emission. GRB pulse durations span five orders of magnitude in the observer's frame, yet time dilation should stretch pulse durations by at most a factor of ten (for $z=9$). Presumably, other intrinsic effects also dominate. For example, relativistic effects caused by the Lorentz factor $\Gamma$ within the GRB jet ({\em e.g.}\ enhanced brightness due to relativistic beaming) can dominate because the blueshift of the jetted material (with $\Gamma$ between $100$ and $1000$) greatly exceeds the cosmological redshift. These effects could explain why bright GRBs tend to be dominated by complex light curves containing multiple, short-duration, short-lag, luminous pulses, while the faintest GRBs more often have long duration, long lag single pulses. Because pulse duration also correlates with fluence, hardness, and anisotropy, each of these observed pulse characteristics are also pulse luminosity indicators.

Our evidence suggests that the pulse peak flux is merely one step in the $E_{\rm peak}$ decay of a pulse spectrum; it does not appear to indicate a phase in the pulse spectral evolution that is different from any other. This is non-intuitive, because peak flux is a pinnacle in the light output, and therefore seems to be a transition between the pulse rise and the pulse decay. However, Figure 10 demonstrates that the pulse peak shifts from high to low energies, and as such is susceptible to the spectral response of the instrument (the instrument cannot measure the early contribution to the pulse peak at energies higher than its upper energy limit, nor can it measure the late contribution at energies lower than its lower energy limit). Despite this limitation, peak flux correlates with all other observable pulse properties. How does the detector sample observable pulse properties so that they always correlate with pulse peak flux, since the value of the pulse peak flux depends on the instrumental response as well as the redshift of the GRB and the intrinsic hardness of the pulse?

This problem is more easily explained by using Occam's Razor to recognize that common (and perhaps ubiquitous) hard-to-soft evolution {\em defines} a pulse better than the increase and subsequent decrease in flux. Once we recognize this, the peak flux is seen just to be a byproduct of the pulse's spectral evolution, rather than its climax. The peak flux therefore represents the predictable sum of flux contributions from different energy ranges produced by the pulse energy decay. {\em By understanding the mechanism capable of producing the pulse spectral evolution, we will understand why all bulk properties of the pulse are correlated.}

\section{The Universality of Long and Short GRB Pulses}

The data presented in Figures 2 and 3 and summarized in Table 2 suggests that the observable properties of GRB pulses, regardless of how the bursts have been classified, make up a single distribution. The observable properties of Long and Short GRBs correlate in similar ways and overlap such that the shortest pulses of Long GRBs have characteristics indistinguishable from typical pulses in Short GRBs. In other words, there seems to be a mechanism capable of producing GRB pulses with correlated properties regardless of the environment in which they form and potentially of the progenitors from which they originate.

The pulse peak flux vs.\ duration relation supports the idea that Long and Short GRB pulses constitute a single distribution, independent of progenitor or environment. A recent analysis of Short GRBs observed by the Swift experiment \citep{nor11} demonstrates that this relation continues to ms timescales (see \cite{nor11} Table 1), and the relation holds regardless of whether the GRB containing them exhibits extended emission (EE) or does not exhibit extended emission (non-EE).

The overlap between Short and Long GRB pulse properties is substantially larger than indicated in Figures 2 and 3 because few of the shortest duration Long GRB pulses have been included in this analysis. These pulses are often found in complex GRBs overlapping other pulses, and thus cannot be fitted easily or uniquely.

Is it possible on the basis of luminosity that Short and Long GRB pulses do not constitute a single distribution? Although other correlated GRB properties appear to be part of a continuous distribution, Short GRBs are known to be less luminous than Long GRBs, suggesting that the peak flux vs.\ duration relation is the result of a serendipitous confluence of conditions such as jet beaming angles, jet Lorentz factors, and burst redshifts that give Short and Long pulses similar peak fluxes when in reality their luminosities are quite disparate. To test this hypothesis, we take a sample containing post-BATSE Short and Long GRBs with known isotropic luminosities \citep{ghi09}. Although we have not explicitly measured the properties of these pulses using our pulse fitting method, we can still roughly estimate the average isotropic luminosity per pulse by dividing the GRB luminosity by the number of pulse peaks. The results, shown in Figure 11, indicate that Short and Long GRB pulses indeed have comparable luminosities, even though they originate from bursts with different luminosities. Long GRBs appear to be more luminous than Short GRBs because they are composed of more pulses having generally larger energy fluences.

It is difficult to accurately measure pulse properties other than peak intensity and duration for Short GRBs due to the short timescale on which these properties are measured. However, pulse peak intensities, durations, and luminosities provide evidence beyond the pulse property correlations {\em suggesting that Short and Long GRB pulses constitute a single phenomenon. If Short and Long GRBs indeed originate from different progenitors, then these pulses suggest that they are a progenitor-independent and environment-independent phenomenon.}

\section{Conclusions}

Pulses are the basic building blocks of GRB prompt emission. However, pulse properties are challenging to measure: they exhibit large individual variations due to intrinsic differences (e.g. intrinsic $E_{\rm peak}$), systematic issues with background rates and instrumental response characteristics, and pervasive pulse overlap. Despite this challenge, all observable GRB pulse properties appear to correlate statistically, with correlated pulse properties being strong in the observer's frame. This surprising result indicates that the inverse-square law and some effects of relativistic cosmology are only of secondary importance to the internal physics responsible for producing the intrinsic pulse properties. GRB bulk properties are constructed by combining and smearing out pulse characteristics in ways that potentially lose valuable information. Since pulse observables have been measured from a four parameter model, it is not surprising that the pulse model parameters themselves are highly correlated. The correlated nature of the observables constrains the underlying physical process capable of producing pulses: the process cannot have too many free parameters because the observed pulse properties might then be largely uncorrelated. Pulses start near-simultaneously in all energy channels and undergo spectral softening after this initial injection. Furthermore, it appears that pulses exhibit remarkably similar temporal evolutions as is indicated by their similar shapes. GRB pulses have progenitor-independent correlated properties: pulse physics appears ubiquitous and is easily replicated across a variety of GRB environments. Since Long GRBs can have high flux, short duration pulses, GRB pulse properties are a red herring in GRB classification; the internal pulse duration {\em distribution} may be a more valuable classification tool. The application of new statistical techniques (T. Loredo, private communication) will allow for better understanding of pulse-measuring limitations in GRBs characterized by considerable pulse overlap.

The correlative properties and ubiquitous nature of GRB pulses places strong constraints on GRB physics. The standard model of GRB prompt emission involves kinematic energy injection via relativistic jets, after which the medium cools. The standard spectral model produces either a synchrotron spectrum \citep{ree94}, or a thermal plus power law spectrum (e.g. \cite{goo86, mes00, dai02, ryd04, ryd05, gui11}). These models, and all other viable models, need to explain the correlated pulse properties and pulse evolution described in this paper. For example, \cite{boc10} point out that the correlative pulse relations cannot be obtained as a direct and simple consequence of the standard synchrotron shock model, since this has no time-dependent component. Many temporal and spectral pulse properties are in rough agreement with predictions related to emitting region curvature within the relativistic jet, although these are model-dependent and rely on offset viewing angle to the jet, time-evolution of the physical process, and other model-dependent effects (e.g. \cite{iok01,lu06,haf07,boc10,ari10}).

\section{Acknowledgments}

We acknowledge the assistance of Alex Greene, Jordan Adams, Charles Nettles, Renate Cumbee, and Jason Ling for analyzing GRB pulse characteristics. We thank Alexander Tchekhovskoy, Demos Kazanas, Jay Norris and Tom Loredo for helpful discussions, and the two journal-supplied referees for their careful and helpful review.  The material presented here is based upon work supported by the NASA Applied Information Systems Research Program under award No.\ NNX09AK60G and through the South Carolina NASA Space Grant program.

\clearpage

\begin{deluxetable}{ccrrrrrrrrcrl}
\tabletypesize{\scriptsize}
\rotate
\tablewidth{0pt}
\tablehead{
\colhead{Pulse ID} & \colhead{class} & \colhead{$w$} & \colhead{$\sigma_w$} & \colhead{$l$} &
\colhead{$\sigma_l$} & \colhead{$p_{\rm 64}$} & \colhead{$\sigma_{\rm p64}$} &
\colhead{HR} & \colhead{$\sigma_{\rm HR}$} &
\colhead{$\kappa$} & \colhead{$S$} & \colhead{$\sigma_S$}
}
\startdata
BATSE 0105 p01 & L & 6.28 & 0.54 & -0.028 & 0.036 & 2.539 & 0.126 & 2.893 & 0.183 & 0.771 & $1.69e-06$ & $4.90e-08$ \\
BATSE 0105 p02 & L & 4.35 & 0.45 & 0.166 & 0.025 & 3.902 & 0.170 & 3.016 & 0.185 & 0.967 & $1.89e-06$ & $5.16e-08$ \\
BATSE 0105 p03 & L & 1.62 & 0.28 & -0.036 & 0.010 & 7.453 & 0.278 & 2.967 & 0.188 & 0.032 & $1.69e-06$ & $4.92e-08$ \\
BATSE 0108 p01 & L & 0.34 & 0.07 & 0.113 & 0.190 & 0.446 & 0.116 & 4.583 & 2.227 & 0.900 & $3.01e-08$ & $6.86e-09$ \\
BATSE 0111 p01 & L & 118.32 & 3.40 & 5.490 & 1.138 & 0.346 & 0.081 & 1.709 & 0.073 & 0.945 & $3.95e-06$ & $7.05e-08$ \\
\enddata
\tablecomments{Electronic Table 1 is published in its entirety in the electronic edition of the {\it Astrophysical Journal}.  A portion is shown here for guidance regarding its form and content. Table 1 contains the pulse properties for 1338 pulses from 610 BATSE GRBs used in this study, along with their uncertainties. These are preliminary data from the BATSE pulse catalog currently under development. Column 1 contains the pulse number (BATSE trigger ID plus fitted pulse number), column 2 contains the burst classification (S=Short and L=Long), columns 3 and 4 contain the duration $w$ and duration uncertainty $\sigma_w$ (measured in s), columns 5 and 6 contain the lag $l$ and lag uncertainty $\sigma_l$ (s), columns 7 and 8 contain the 64 ms peak flux $p_{\rm 64}$ and peak flux uncertainty $\sigma_{\rm p64}$ (counts ${\rm cm^{-2} s^{-1}}$), columns 9 and 10 contain the energy hardness ratio HR and HR uncertainty $\sigma_{\rm HR}$, column 11 contains the asymmetry $\kappa$ (uncertainties $\sigma_\kappa$ all assumed to be $\pm 0.05$), and columns 12 and 13 contain the channel $1+2+3$ energy fluence $S$ and fluence uncertainty $\sigma_S$ (ergs ${\rm cm^{-2} s^{-1}}$).}
\end{deluxetable}

\clearpage

\begin{table}
\begin{tabular}{lccccc}
\hline
\multicolumn{6}{c}{Pulses from Long GRBs} \\
\hline
 & lag & peak flux & hardness & asymmetry & fluence \\
\hline
duration & $\bf{< 0.01\%}$ & $\bf{< 0.01\%}^\dagger$ & $\bf{< 0.01\%}^\dagger$ & $\bf{< 0.01\%}$ & $\bf{< 0.01\%}$\\
lag & --- & $\bf{< 0.01\%^\dagger}$ & $\bf{< 0.01\%}^\dagger $ & $\bf{< 0.01\%}$ & $\bf{< 0.01\%}$\\
peak flux & --- & --- & $\bf{< 0.01\%}$ & $\bf{0.03\%}^\dagger$ & $\bf{< 0.01\%}$\\
hardness & --- & --- & --- & $\bf{< 0.01\%}^\dagger$ & $3.1\%^\dagger $\\
asymmetry & --- & --- & --- & --- & $\bf{< 0.01\%}$\\
\hline
\multicolumn{6}{c}{Pulses from Short GRBs} \\
\hline
 & lag & peak flux & hardness & asymmetry & fluence \\
\hline
duration &$\bf{< 0.01\%}$ & $\bf{< 0.01\%}^\dagger$ & $6.8\%^\dagger$ & $76\%^\dagger$ & $11\%^\dagger$\\
lag & --- & $54\%^\dagger$ & $25\%^\dagger$ & $22\%$ & $50\%$\\
peak flux & --- & --- & $\bf{< 0.01\%}$ & $54\%$ & $\bf{< 0.01\%}$\\
hardness & --- & --- & --- & $93\%^\dagger$ & $\bf{< 0.01\%}$\\
asymmetry & --- & --- & --- & --- & $43\%$\\
\hline

\end{tabular}
\caption{GRB pulse correlations. Spearman rank order correlation probabilities that the pulse characteristics in question are as expected from random, uncorrelated distributions. Anti-correlations are indicated by a $\dagger$, and boldface numbers indicate statistically significant correlation/anti-correlations.}
\label{Table 2}
\end{table}

\clearpage

\begin{table}
\begin{tabular}{lccccc}
\hline
\multicolumn{6}{c}{Percentage of Multi-Pulsed GRBs Showing Overall Pulse Behavior} \\
\hline
 & lag & peak flux & hardness & asymmetry & fluence \\
\hline
duration & $0.64\%$ & $0.63\%^\dagger$ & $0.61\%^\dagger$ & $0.67\%$ & $0.77\%$\\
lag & --- & $0.50\%^\dagger$ & $0.48\%^\dagger $ & $0.59\%$ & $0.65\%$\\
peak flux & --- & --- & $0.69\%$ & $0.54\%^\dagger$ & $0.72\%$\\
hardness & --- & --- & --- & $0.57\%^\dagger$ & $0.59\%^\dagger $\\
asymmetry & --- & --- & --- & --- & $0.59\%$\\
\hline

\end{tabular}
\caption{Multi-pulsed Short and Long GRBs have been combined together to provide a larger sample; least squares is applied to the properties of all pulses fitted within each multi-pulsed GRB to learn more about the intra-pulse distribution. The resultant gradient (slope) of each pulse property relative to the others indicates whether the properties within the burst are correlated (positive gradient) or anti-correlated (negative gradient; indicated by a $\dagger$). The percentage of multi-pulsed GRBs exhibiting correlated behaviors is then compared to the results of the inter-pulse duration shown in Figure 2 and Table 2; if somewhat more than 50\% of the multi-pulsed GRBs have the same general gradient as that obtained from the inter-pulse distribution then the two distributions have the same general behavior. Since these percentages average 62\%, the behaviors of the intra-pulse distribution have the same sense but otherwise appear to be generally weaker than the power-law forms of the inter-pulse distribution.}
\label{Table 3}
\end{table}

\clearpage

\begin{table}
\begin{tabular}{lcccc}
\hline
BATSE Trigger & N pulses & Classification (Source) & Disposition & Explanation \\
\hline
563 & 1 & H2S (P1) & keep & clean fit \\
647 & $>5$ & H2S (P1) & discard & overlapping pulses\\
1156 & n & IT (P2) & discard & overlapping pulses\\
1406 & 1 & H2S (P1) & keep & clean fit\\
1733 & 1 & IT (L1, P2) & reconsider & multiple pulses? \\
1883 & 1 & H2S (L1, P1) & keep & clean fit\\
1956 & 1 & IT (L1) \& H2S (P1) & exclude & dual classification\\
2083 p. 1 & 2 & IT (P2, L1) \& H2S (P1) & exclude & dual classification\\
2083 p. 2 & 2 & IT (P2, L1) \& H2S (P1) & exclude & dual classification\\
2138 p. 1 & n & IT (P1) \& H2S (P2) & exclude & dual classification\\
2138 p. 2 & n & IT (P1) \& H2S (P2) & exclude & dual classification\\
2193 & 1 & H2S (L1, P1) & keep & clean fit\\
2387 & 1 & H2S (L1, P1) & keep & clean fit\\
2389 & $>1$ & IT (P2) & discard & overlapping pulses\\
2519 & 2 & H2S (P1) & keep & clean fit\\
2662 & 1 & H2S (P1) & reconsider & multiple pulses?\\
2665 & 1 & H2S (P1) & reconsider & multiple pulses?\\
2700 & n & H2S (P1) & discard & overlapping pulses\\
2812 & n & IT (P2) & reconsider & multiple pulses?\\
2880 & 1 & H2S (P1) & keep & clean fit\\
2919 & $>1$ & IT (P2) & discard & overlapping pulses\\
3003 & 1 & IT (L1, P2) \& H2S (P1) & exclude & dual classification\\
\hline
\end{tabular}
\caption{Hard-to-soft (H2S) and Intensity tracking (IT) pulses in common with BATSE pulse database. Pulse classification sources are P1=\citep{pen09a}, P2=\citep{pen09b}, L1=\citep{lu10}. }
\label{Table 4}
\end{table}

\clearpage

\begin{figure}
\plottwo{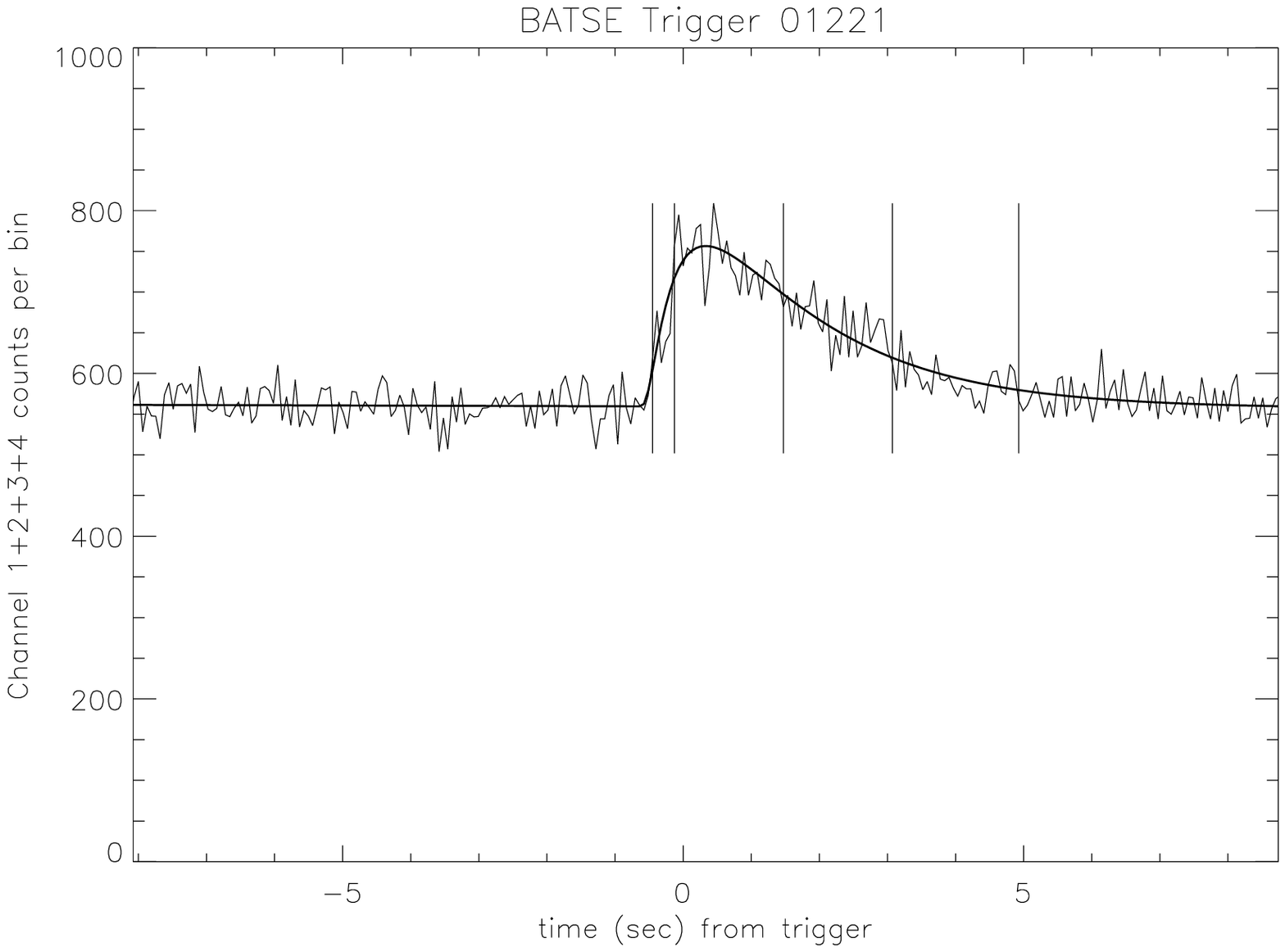} {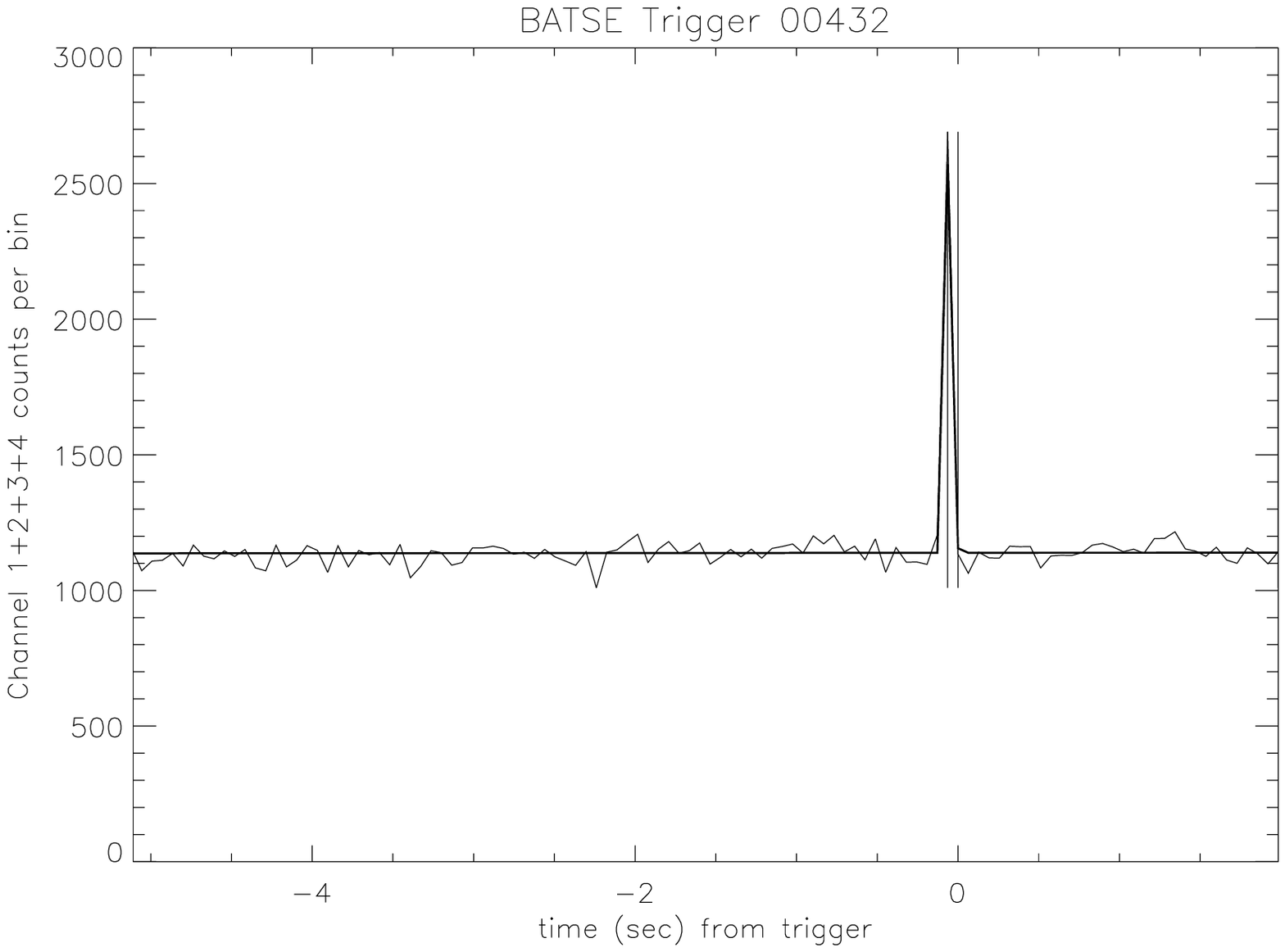}
\plottwo{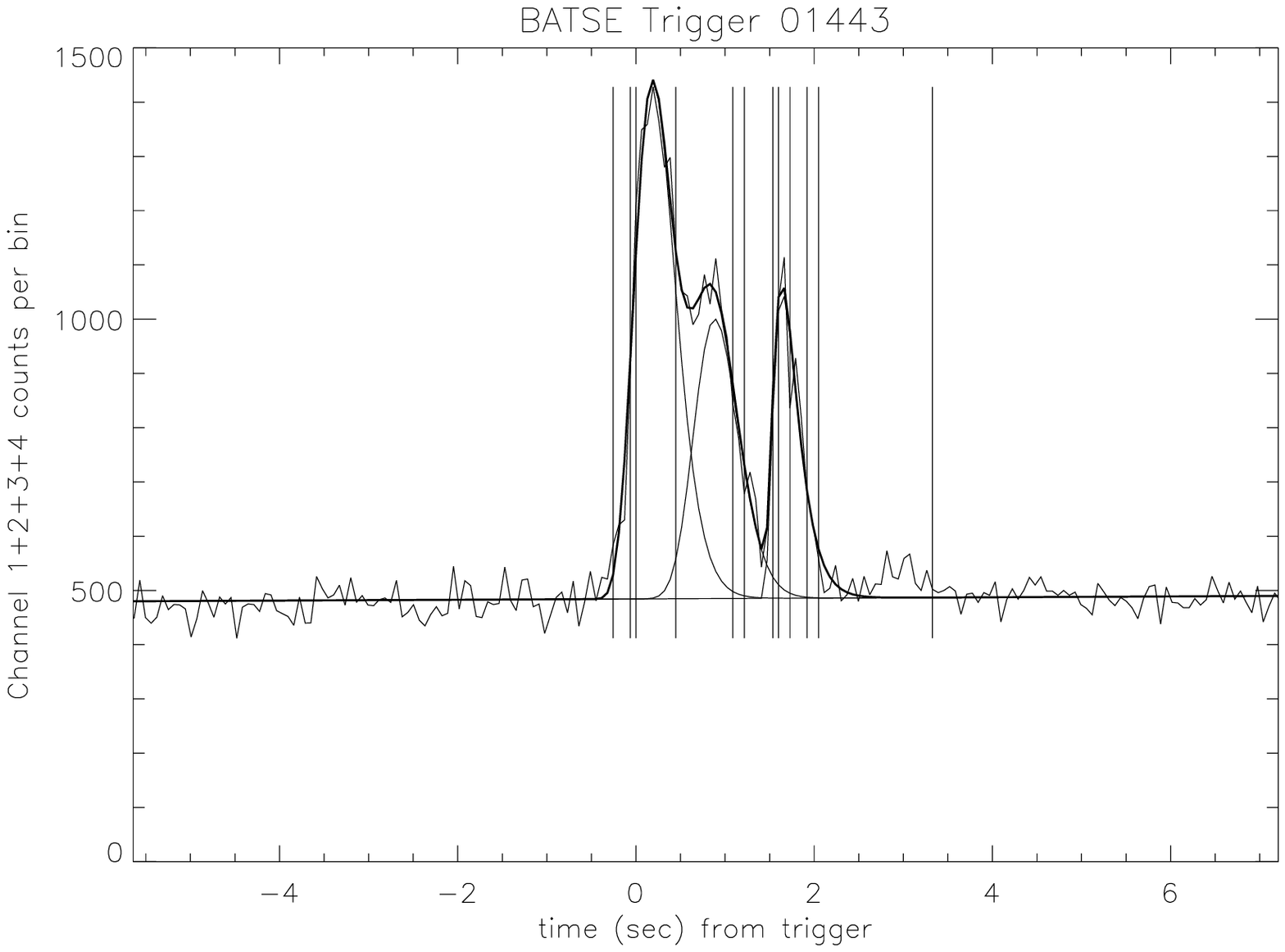} {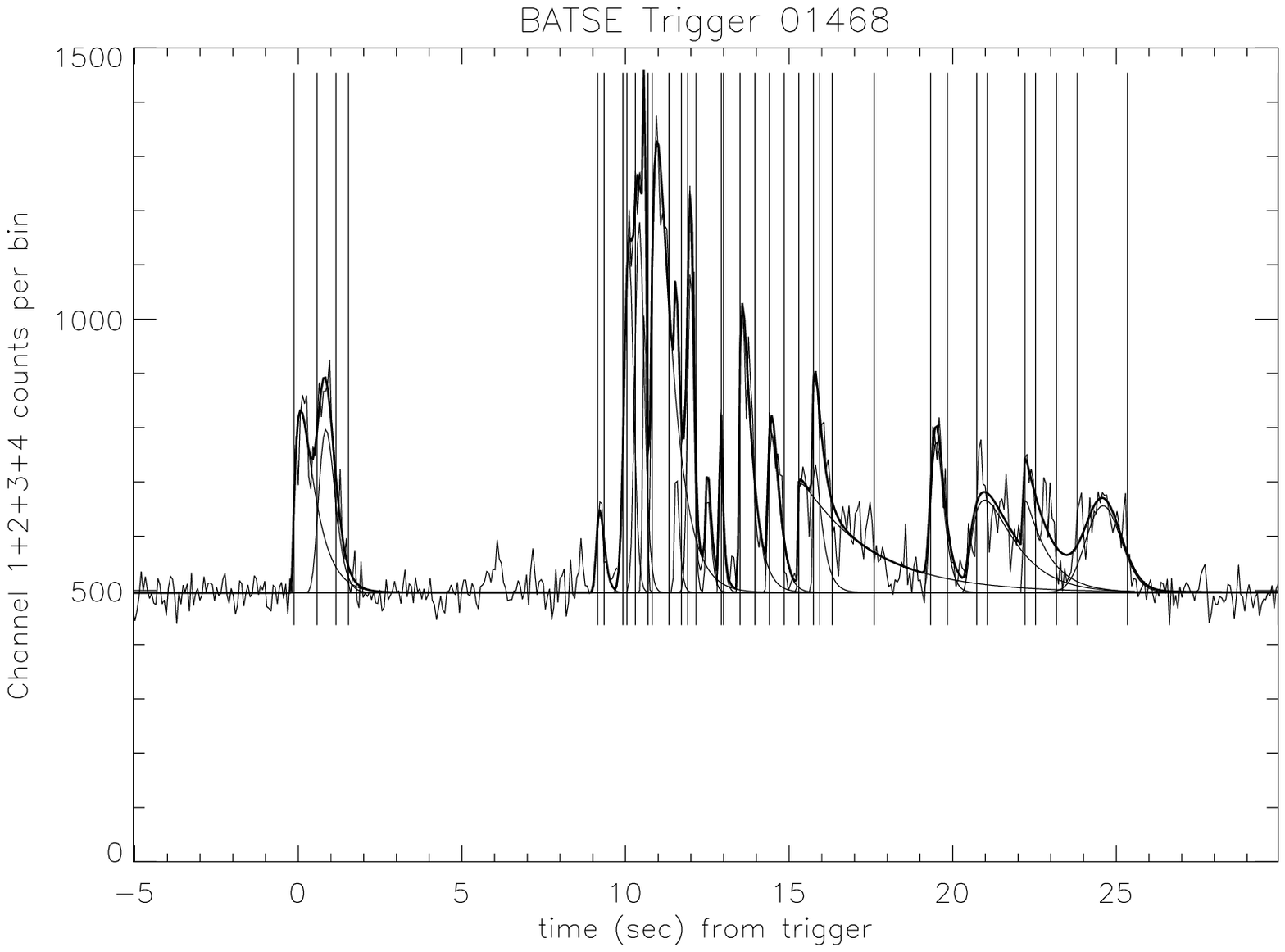}
  \caption{Sample 4-channel pulse fits for BATSE triggers 1221 (upper left), 432 (upper right), 1443 (lower left) and 1468 (lower right). Overlaying the 64-ms data are vertical lines defining each Bayesian block segment, as well as the pulses finally fitted to the data.}
\end{figure}

\clearpage

\begin{figure}
  \includegraphics{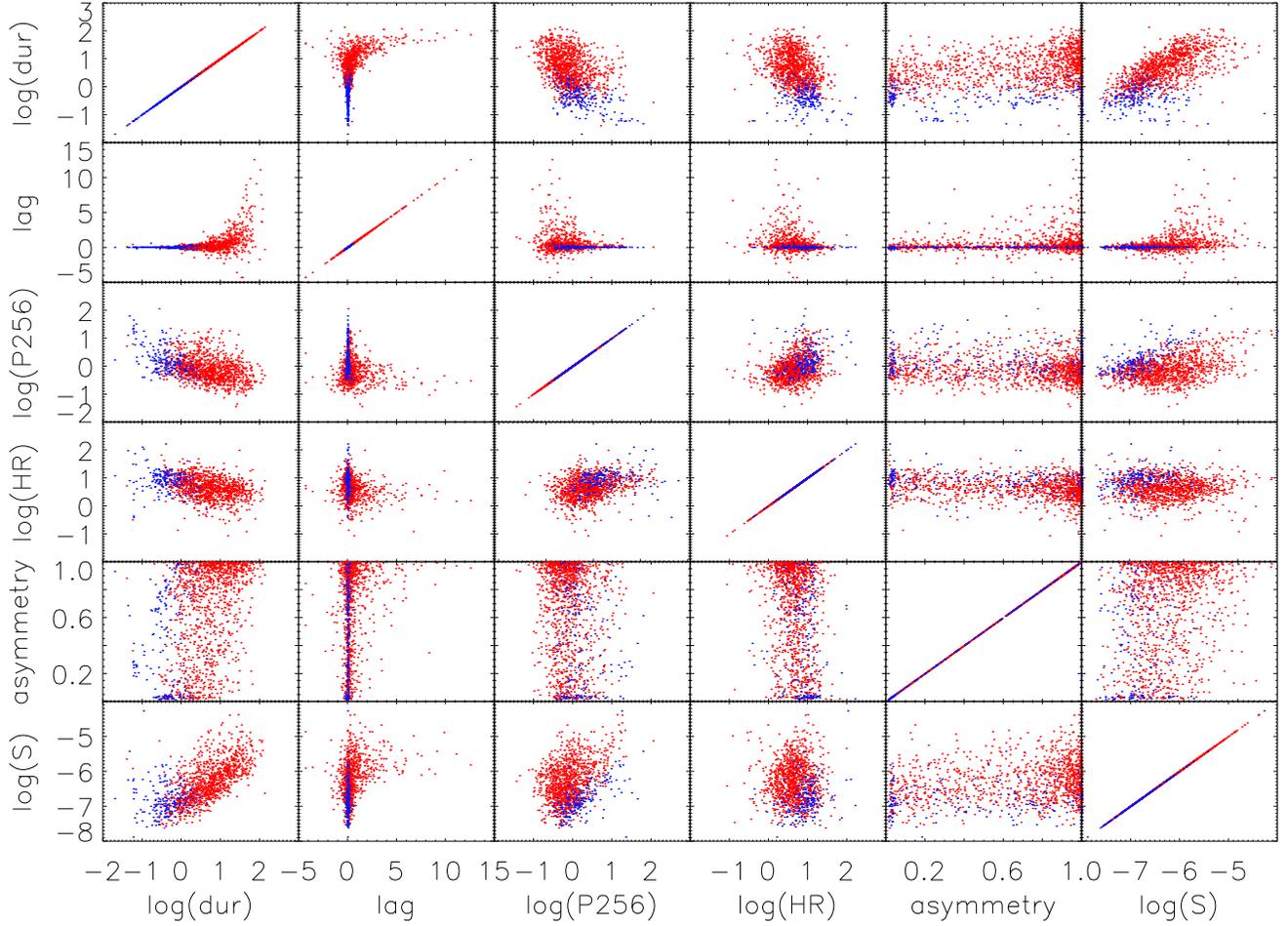}
  \caption{Pulse correlations for Long (red) and Short (blue) GRB pulses, showing the strong correlation between these parameters.}
\end{figure}

\clearpage

\begin{figure}
  \includegraphics{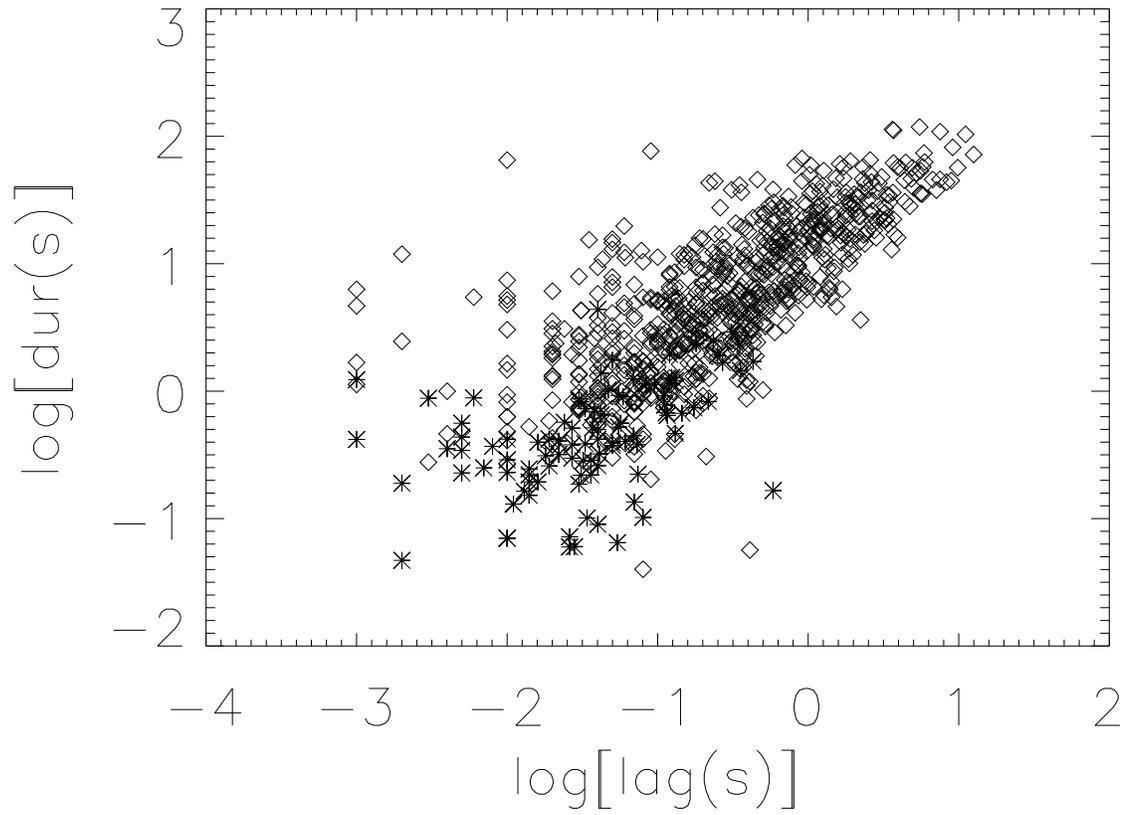}
  \caption{Pulse duration vs.\ pulse peak lag for Long (diamond) and Short (*) GRB pulses, showing the strong correlation between these parameters. Only pulses with positive lags are plotted.}
\end{figure}

\clearpage

\begin{figure}
  \includegraphics{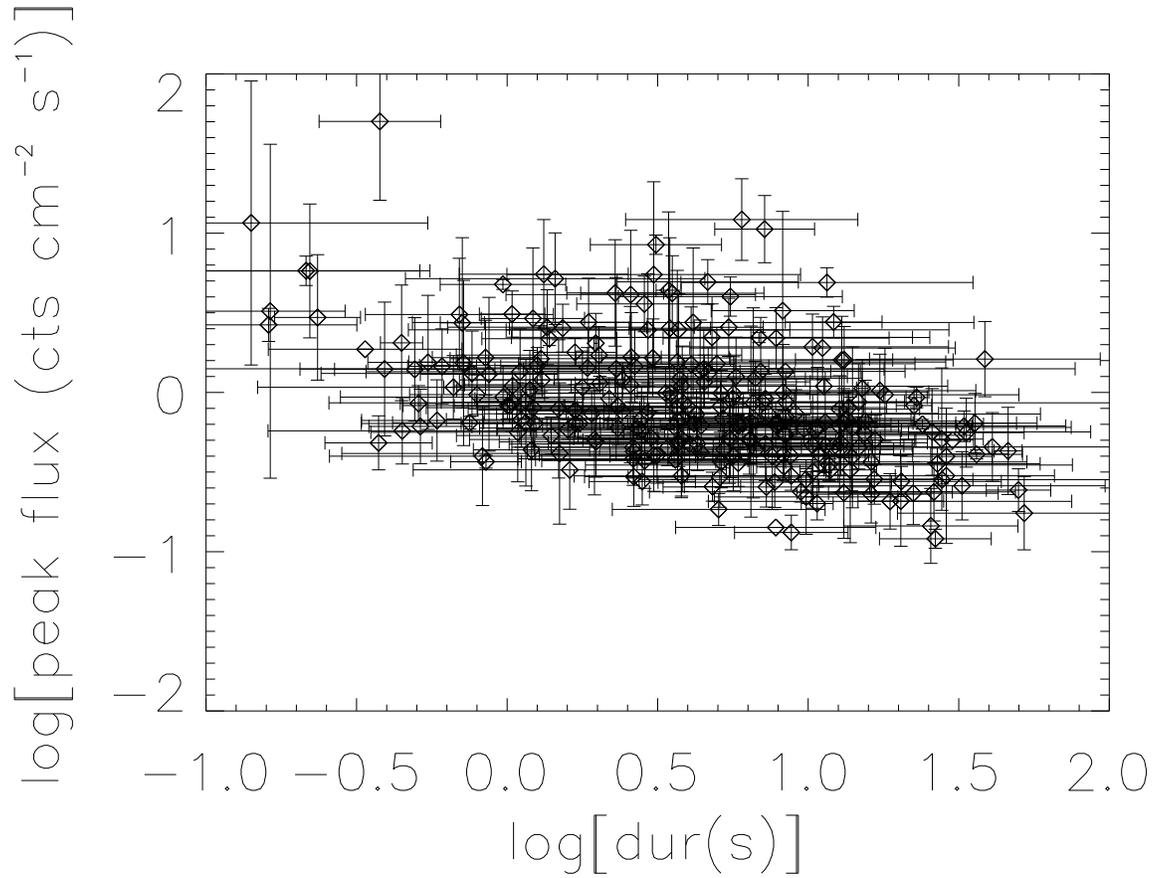}
  \caption{Pulse 256 ms peak flux vs.\ duration for multi-pulsed GRBs, indicating general intra-pulse characteristics. Each data point identifies the mean peak flux vs.\ mean duration for a mutli-pulsed GRB; the error bars reflect the mean error. The shorter the average pulse duration is in a burst, the brighter the average pulse is.}
\end{figure}

\clearpage

\begin{figure}
  \includegraphics{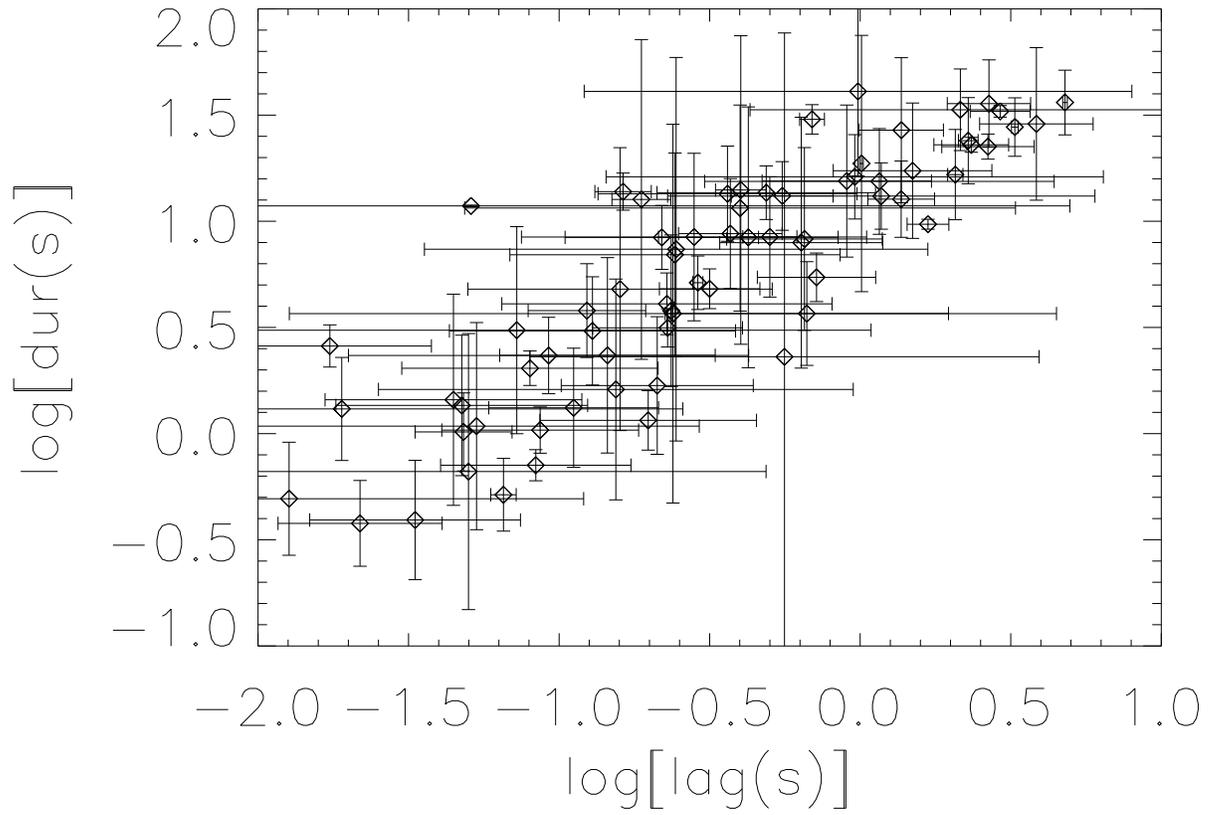}
  \caption{Pulse duration vs.\ lag for multi-pulsed GRBs using the same notation as Figure 4 (only bursts with positive mean pulse lags are plotted). The greater the average pulse duration is in a burst, the longer average lag the pulse will have.}
\end{figure}

\clearpage

\begin{figure}
  \includegraphics{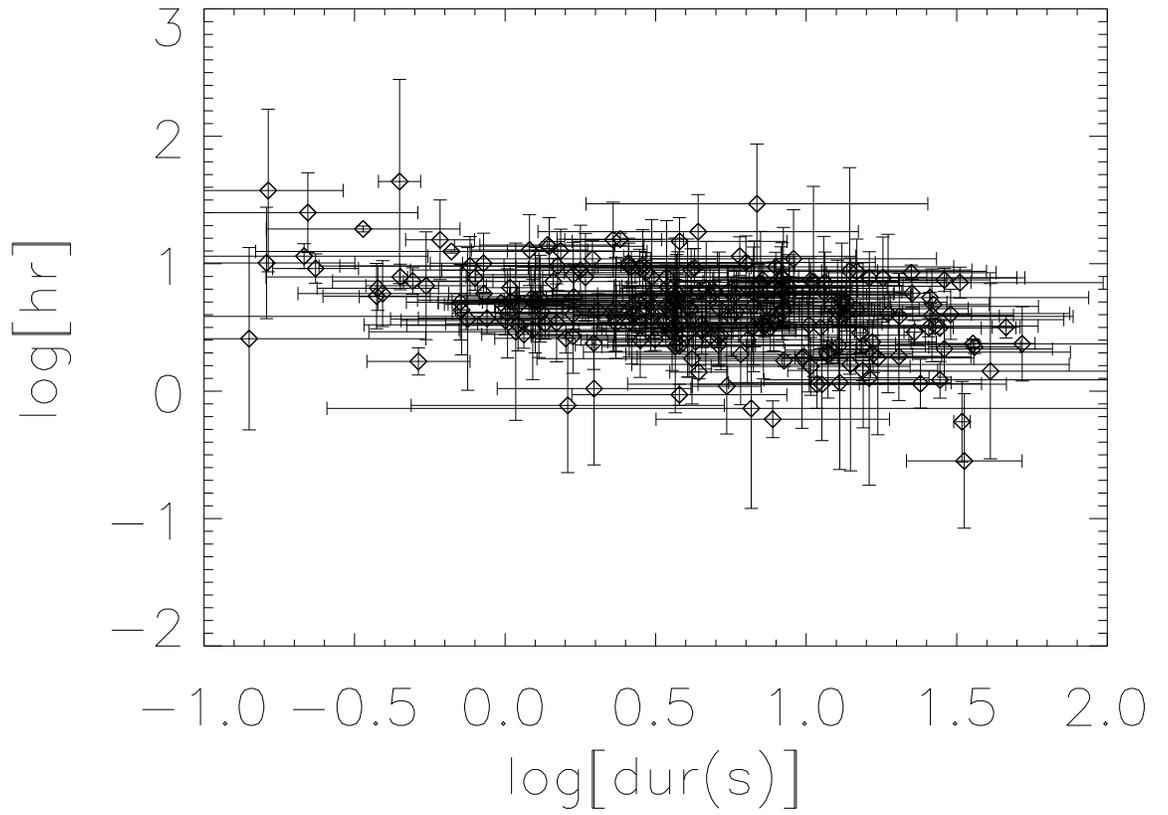}
  \caption{Pulse hardness vs.\ duration for multi-pulsed GRBs, using the same notation as Figure 4. The harder the average pulse is for a multi-pulsed GRB, the shorter the average pulse will be.}
\end{figure}

\clearpage

\begin{figure}
  \includegraphics{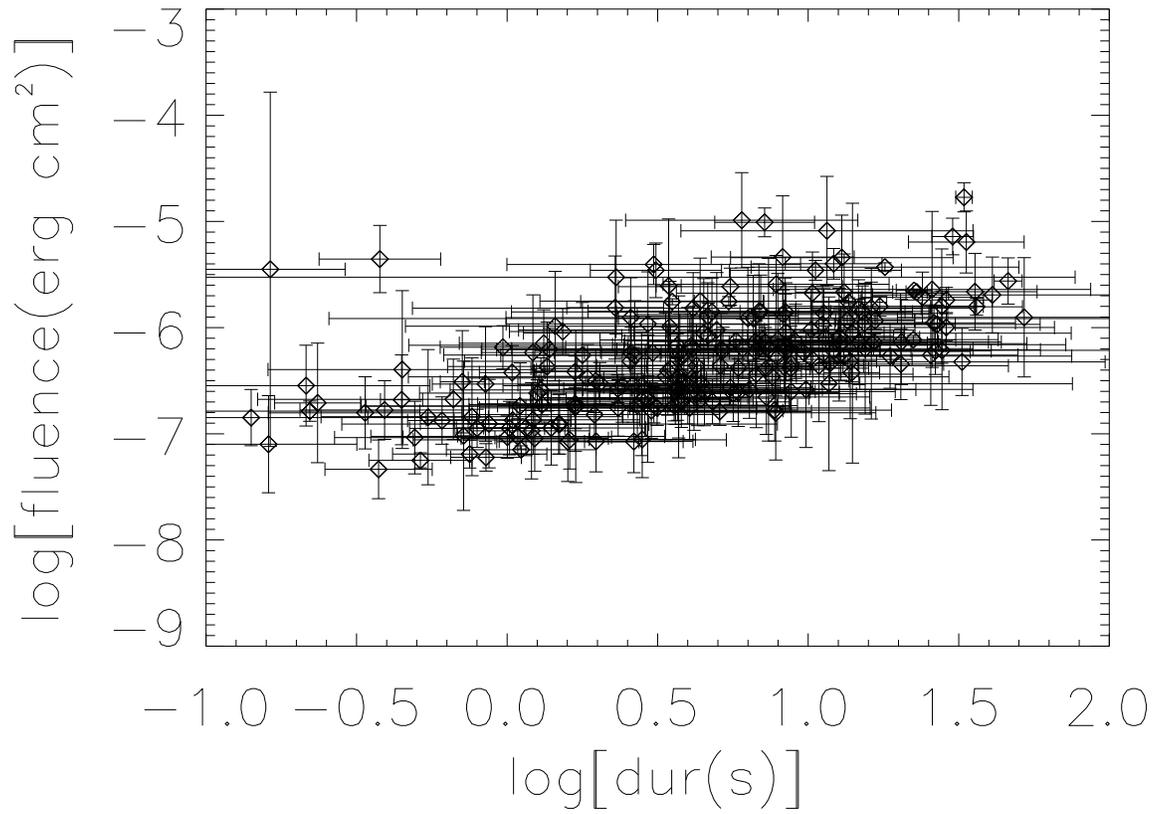}
  \caption{Pulse fluence vs.\ duration for multi-pulsed GRBs, using the same notation as Figure 4. The greater average energy fluence a pulse has, the longer the average pulse will be.}
\end{figure}

\clearpage

\begin{figure}
  \includegraphics{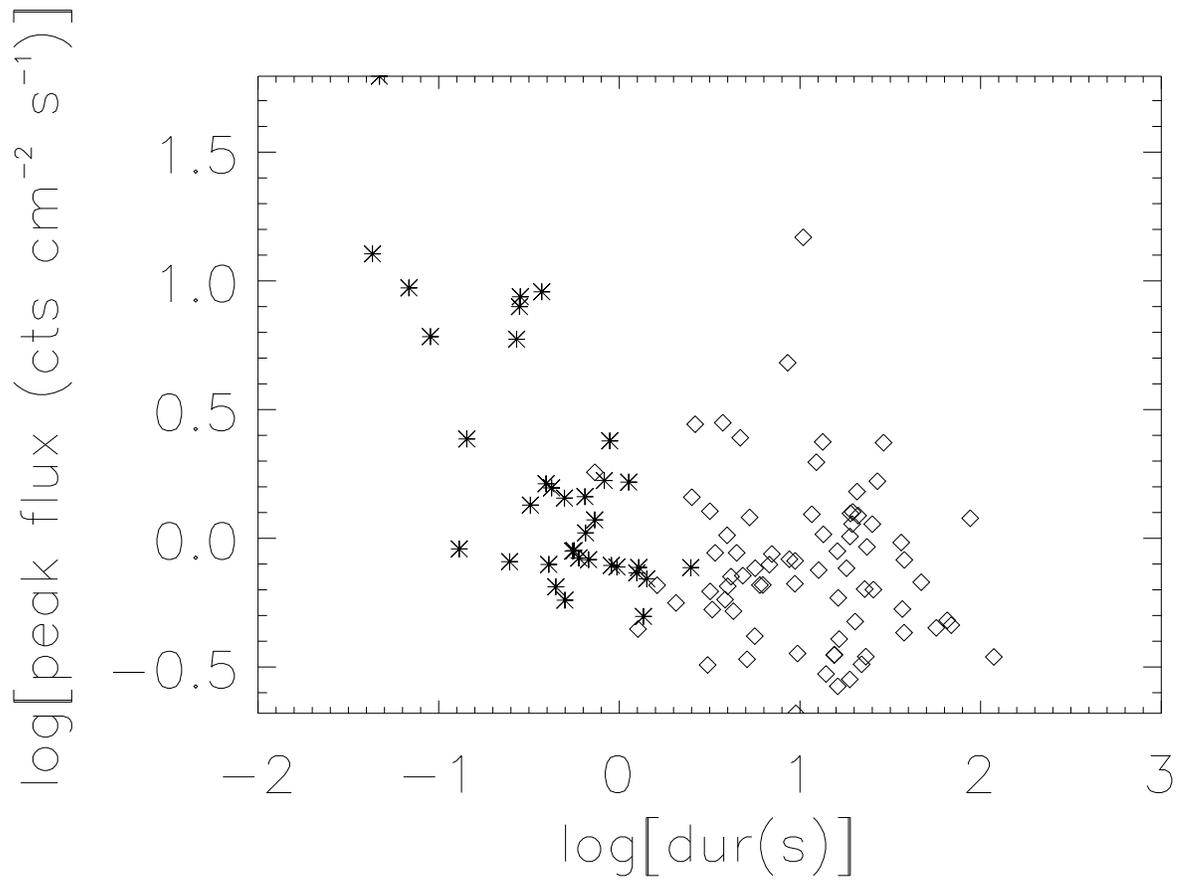}
  \caption{Pulse 256 ms peak flux vs.\ duration for Long (diamond) and Short (*) GRB pulses for single-pulsed GRBs triggering on the 256 ms timescale.}
\end{figure}

\clearpage

\begin{figure}
  \includegraphics{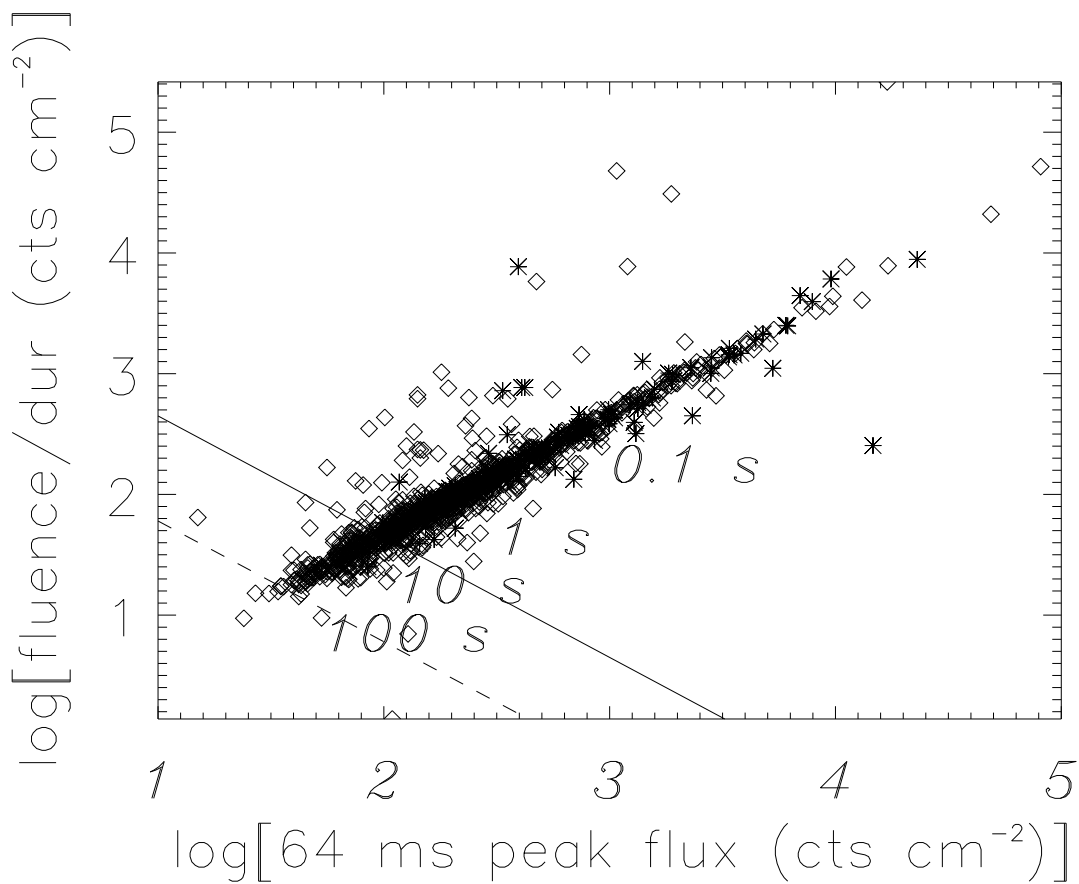}
  \caption{Estimated limitations on pulse sampling threshold for Long (diamond) and Short (*) GRB pulses in terms of fluence/duration and 64 ms peak flux. Sampling is relatively complete above and to the right of the solid line, with the noted exception of pulses in fields too crowded to cleanly fit. The estimated lower limit (below which almost no pulses have been fitted) is indicated by the dotted line. Approximate durations of the plotted pulses are given.}
\end{figure}

\clearpage

\begin{figure}
  \includegraphics{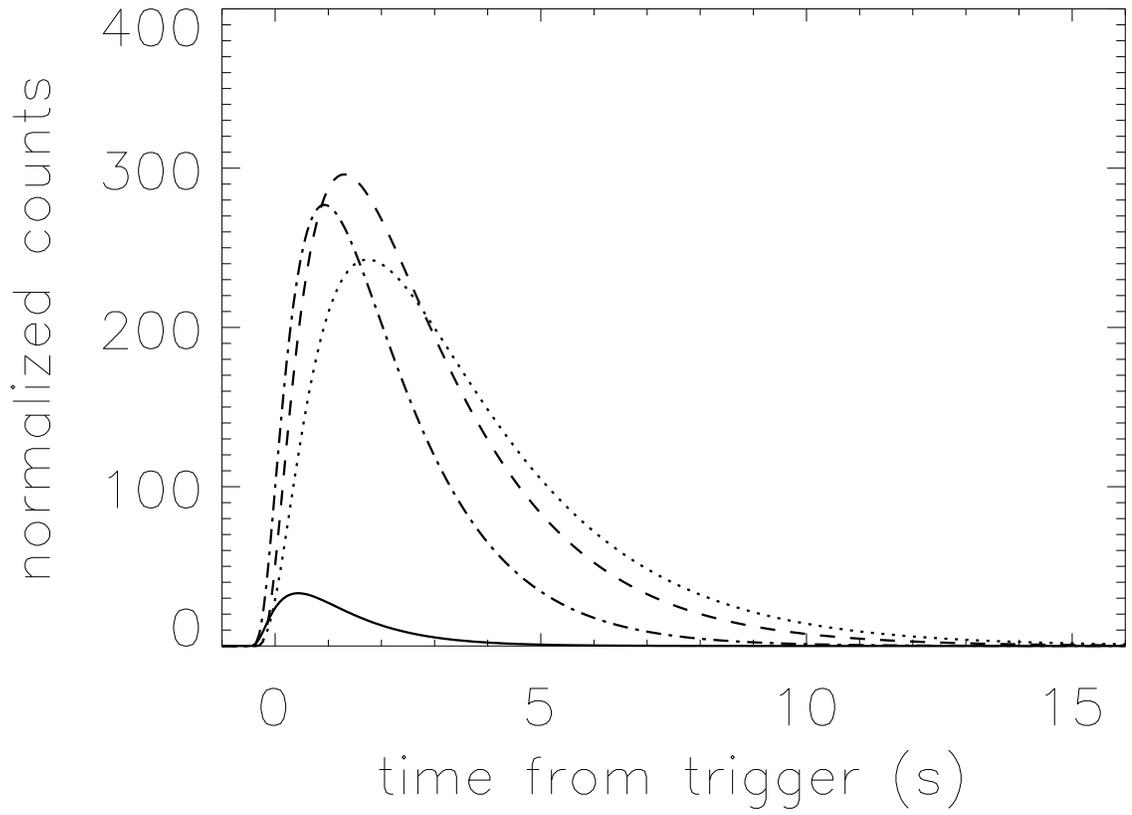}
  \caption{GRB pulse spectral evolution, as seen from the fits to single-pulsed BATSE Trigger 1883. The solid line denotes the channel 4 fit, the dotted-dashed line is channel 3 fit, the dashed line is channel 2 fit, and the dotted line is channel 1 fit.}
\end{figure}

\clearpage

\begin{figure}
  \includegraphics{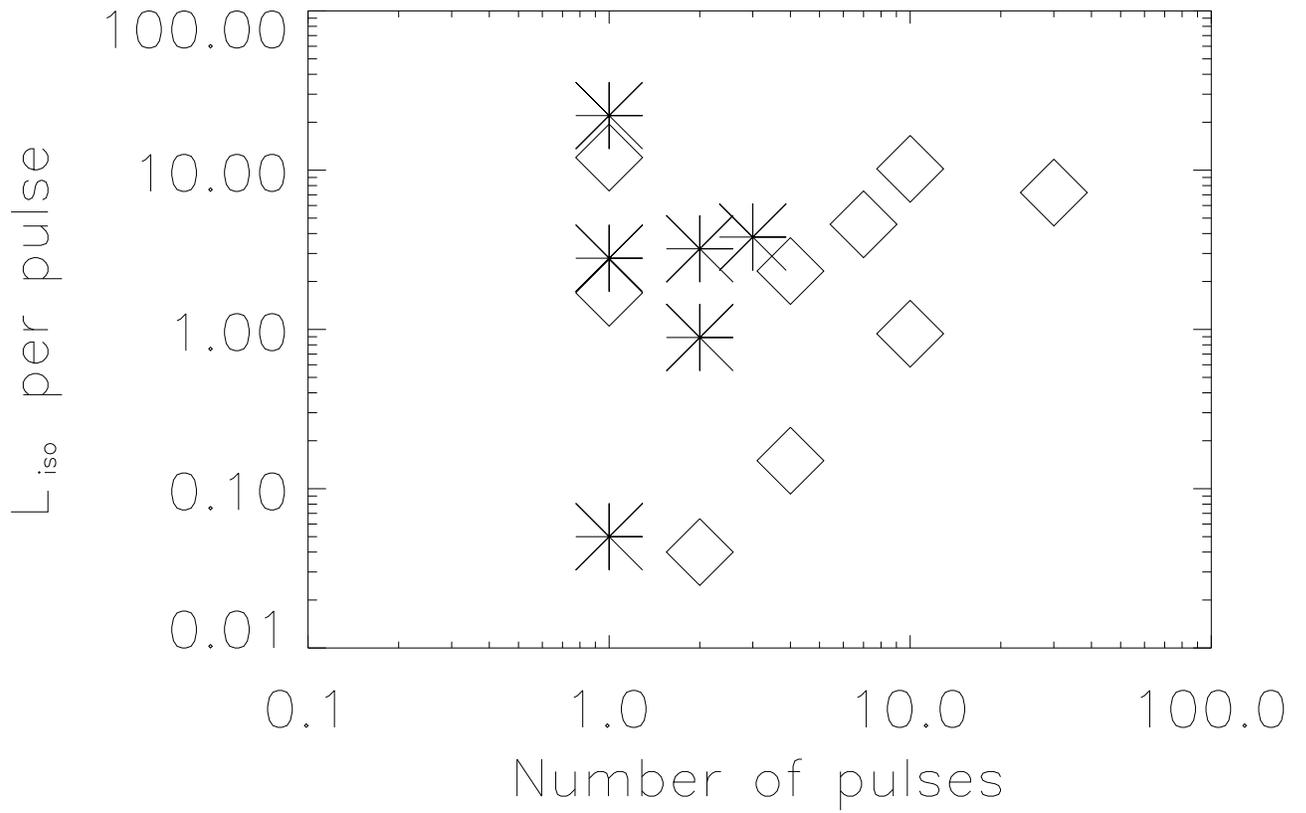}
  \caption{Approximate isotropic luminosity per pulse for Long (diamond) and Short (*) GRB pulses (data from \cite{ghi09}), demonstrating that Short GRB pulses are not systematically less luminous than Long GRB pulses, even though Short GRBs are less luminous than Long GRBs.}
\end{figure}


\begin{thebibliography}{}
\bibitem[Acton(1970)]{act70} Acton, F.~S.\ 1970, ``Numerical Methods that Work'' (Harper and Row, NY).
\bibitem[Amati et al.(2002)]{ama02} Amati, L., et al.\ 2002, \aap, 390, 81
\bibitem[Arimoto et al.(2010)]{ari10} Arimoto, M., et al.\ 2010, \pasj, in press
\bibitem[Band et al.(1993)]{ban93} Band, D., et al.\ 1993, \apj, 413, 281
\bibitem[Band(1997)]{ban97} Band, D.~L.\ 1997, \apj, 486, 928
\bibitem[Bo{\c c}i et al.(2010)]{boc10} Bo{\c c}i, S., Hafizi, M., \& Mochkovitch, R.\ 2010, \aap, 519, A76 
\bibitem[Chincarini et al.(2010)]{chi10} Chincarini, G., et al.\ 2010, \mnras, 406, 2113 
\bibitem[Crider et al.(1999)]{cri99} Crider, A., Liang, E.~P., Preece, R.~D., Briggs, M.~S., Pendleton, G.~N., Paciesas, W.~S., Band, D.~L., \& Matteson, J.~L.\ 1999, \aaps, 138, 401 
\bibitem[Daigne \& Mochkovitch(2002)]{dai02} Daigne, F., \& Mochkovitch, R.\ 2002, \mnras, 336, 1271
\bibitem[Ford et al.(1995)]{for95} Ford, L.~A., et al.\ 1995, \apj, 439, 307 
\bibitem[Ghirlanda et al.(2009)]{ghi09} Ghirlanda, G., Nava, L., Ghisellini, G., Celotti, A., \& Firmani, C.\ 2009, \aap, 496, 585 
\bibitem[Golenetskii et al.(1983)]{gol83} Golenetskii, S.~V., Mazets, E.~P., Aptekar, R.~L., \& Ilinskii, V.~N.\ 1983, \nat, 306, 451 
\bibitem[Goodman(1986)]{goo86} Goodman, J.\ 1986, \apjl, 308, L47
\bibitem[Guiriec et al.(2011)]{gui11} Guiriec, S., et al.\ 2011, \apjl, 727, L33 
\bibitem[Hafizi \& Mochkovitch(2007)]{haf07} Hafizi, M., \& Mochkovitch, R.\ 2007, \aap, 465, 67
\bibitem[Hakkila et al.(2003)]{hak03} Hakkila, J., Giblin, T.~W., Roiger, R.~J., Haglin, D.~J., 
Paciesas, W.~S., \& Meegan, C.~A.\ 2003, \apj, 582, 320 
\bibitem[Hakkila et al.(2007)]{hak07} Hakkila, J., et al.\ 2007, \apjs, 169, 62
\bibitem[Hakkila et al.(2008)]{hak08} Hakkila, J., et al. \ 2008a, \apjl, 677, L81
\bibitem[Hakkila \& Cumbee(2009a)]{hak09a} Hakkila, J., \& Cumbee, R.~S.\ 2009, in AIP Proc. 1133 (ed. Meegan, Gehrels, \& Kouveliotou), 379
\bibitem[Hakkila, Fragile, and Giblin(2009b)]{hak09b} Hakkila, J., Fragile, P.~C., \& Giblin, T.~W.\ 2009, in AIP Proc. 1133 (ed. Meegan, Gehrels, \& Kouveliotou), 479
\bibitem[Hakkila \& Nemiroff(2009)]{hak09c} Hakkila, J., \& Nemiroff, R.~J.\ 2009, \apj, 705, 372 
\bibitem[Higdon \& Lingenfelter(1996)]{hig96} Higdon, J.~C., \& Lingenfelter, R.~E.\ 1996, American Institute of Physics Conference Series, 384, 402 
\bibitem[Horack \& Hakkila(1997)]{hor97} Horack, J.~M., \& Hakkila, J.\ 1997, \apj, 479, 371 
\bibitem[Horv{\'a}th(1998)]{hor98} Horv{\'a}th, I.\ 1998, \apj, 508, 757 
\bibitem[Ioka \& Nakamura(2001)]{iok01} Ioka, K., \& Nakamura, T.\ 2001, \apjl, 554, L163 
\bibitem[Kaneko et al.(2006)]{kan06} Kaneko, Y., Preece, R.~D., Briggs, M.~S., Paciesas, W.~S., Meegan, C.~A., \& Band, D.~L.\ 2006, \apjs, 166, 298 

\bibitem[Kocevski et al.(2003)]{koc03} Kocevski, D., Ryde, F., \& Liang, E.\ 2003, \apj, 596, 389 

\bibitem[Kouveliotou et al.(1993)]{kou93} Kouveliotou, C., Meegan, C.~A., Fishman, G.~J., Bhat, N.~P., Briggs, M.~S., Koshut, T.~M., Paciesas, W.~S., \& Pendleton, G.~N.\ 1993, \apjl, 413, L101 
\bibitem[Liang \& Kargatis(1996)]{lia96} Liang, E., \& Kargatis, V.\ 1996, \nat, 381, 49 
\bibitem[Lu et al.(2006)]{lu06} Lu, R.-J., Qin, Y.-P., Zhang, Z.-B., \& Yi, T.-F.\ 2006, \mnras, 367, 275 
\bibitem[Lu et al.(2010)]{lu10} Lu, R.-J., Hou, S.-J., \& Liang, E.-W.\ 2010, \apj, 720, 1146 
\bibitem[Margutti et al.(2010)]{mag10} Margutti, R., Guidorzi, C., Chincarini, G., Bernardini, M.~G., Genet, F., Mao, J., \& Pasotti, F.\ 2010, \mnras, 406, 2149 
\bibitem[Meegan et al.(2000)]{mee00} Meegan, C., Hakkila, J., Johnson, A., Pendleton, G., \& Mallozzi, R.\ 2000, Gamma-ray Bursts, 5th Huntsville Symposium, 526, 43
\bibitem[M{\'e}sz{\'a}ros \& Rees(2000)]{mes00} M{\'e}sz{\'a}ros, P., \& Rees, M.~J.\ 2000, \apj, 530, 292
\bibitem[M{\'e}sz{\'a}ros et al.(2011)]{mes11} M{\'e}sz{\'a}ros, A., {\v R}{\'{\i}}pa, J., \& Ryde, F.\ 2011, \aap, 529, A55 
\bibitem[Mukherjee et al.(1998)]{muk98} Mukherjee~S., et al. 1998, \apj, 508, 314
\bibitem[Nemiroff(2000)]{nem00} Nemiroff, R.~J.\ 2000, \apj, 544, 805 
\bibitem[Norris et al.(1996)]{nor96} Norris, J.~P., Nemiroff, R.~J., Bonnell, J.~T., Scargle, 
J.~D., Kouveliotou, C., Paciesas, W.~S., Meegan, C.~A., \& Fishman, G.~J.\ 1996, \apj, 459, 393 
\bibitem[Norris et al.(2000)]{nor00} Norris, J.~P., Marani, G.~F., \& Bonnell, J.~T.\ 2000, \apj, 534, 248 
\bibitem[Norris(2002)]{nor02} Norris, J.~P.\ 2002, \apj, 579, 386 
\bibitem[Norris et al.(2005)]{nor05} Norris, J.~P., Bonnell, J.~T., Kazanas, D., Scargle, J.~D., Hakkila, J., \& Giblin, T.~W.\ 2005, \apj, 627, 324 
\bibitem[Norris et al.(2011)]{nor11} Norris, J.~P., Gehrels, N., \& Scargle, J.~D.\ 2011, arXiv:1101.1648 
\bibitem[Peng et al.(2009a)]{pen09a} Peng, Z.~Y., Ma, L., Zhao, X.~H., Yin, Y., Fang, L.~M., \& Bao, Y.~Y.\ 2009, \apj, 698, 417 
\bibitem[Peng et al.(2009b)]{pen09b} Peng, Z.~Y., Ma, L., Lu, R.~J., Fang, L.~M., Bao, Y.~Y., \& Yin, Y.\ 2009, \na, 14, 311
\bibitem[Piran(2005)]{pir05} Piran, T.\ 2005, Reviews of Modern Physics, 76, 1143 
\bibitem[Ramirez-Ruiz \& Fenimore(2000)]{rrm00} Ramirez-Ruiz, E., \& Fenimore, E.~E.\ 2000, \apj, 539, 712
\bibitem[Rees \& Meszaros(1994)]{ree94} Rees, M.~J., \& Meszaros, P.\ 1994, \apjl, 430, L93 
\bibitem[Reichart et al.(2001)]{rei01} Reichart, D.~E., Lamb, D.~Q., Fenimore, E.~E., Ramirez-Ruiz, E., Cline, T.~L., \& Hurley, K.\ 2001, \apj, 552, 57 
\bibitem[Richardson et al.(1996)]{ric96} Richardson, G., Koshut, T., Paciesas, W., 
\& Kouveliotou, C.\ 1996, American Institute of Physics Conference Series, 384, 87 
\bibitem[Ryde(2004)]{ryd04} Ryde, F.\ 2004, \apj, 614, 827
\bibitem[Ryde(2005)]{ryd05} Ryde, F.\ 2005, \aap, 429, 869
\bibitem[Scargle(1998)]{sca98} Scargle, J.~D.\ 1998, \apj, 504, 405
\end{thebibliography}
\end{document}